\documentclass[12pt]{article}

\pdfoutput=1

\usepackage{natbib,amsfonts}
\usepackage{amsfonts}
\usepackage{bm}
\usepackage{amsthm}
\usepackage{amsmath}
\usepackage[mathscr]{eucal}
\usepackage[final]{graphicx}

\usepackage{verbatim}

\begin{document}

\setlength{\textwidth}{6.25in}
\setlength{\textheight}{8.5in}
\setlength{\oddsidemargin}{.5in}
\setlength{\topmargin}{-.25in}


\newcommand{\sep}{\vspace{2mm}}
\newcommand{\BR}{\mathbb{R}}
\newcommand{\beq}{\begin{equation}}
\newcommand{\eeq}{\end{equation}}
\newcommand{\smd}{\mathscr{D}}
\newcommand{\eqdef}{\overset{\text{\rm\small def}}=}
\newcommand{\smm}{\mathscr{M}}
\newcommand{\smn}{\mathscr{N}}
\newcommand{\fma}{\mathfrak{A}}
\newcommand{\BP}{\mathbb{P}\,}
\newcommand{\fmm}{\mathfrak{M}}
\newcommand{\BI}{\mathbb{I}}
\newcommand{\BB}{\mathbb{B}}
\newcommand{\BQ}{\mathbb{Q}}
\newcommand{\be}{\begin{equation}}
\newcommand{\ee}{\end{equation}}
\newcommand{\bx}{{\bf x}}
\newcommand{\br}{{\bf y}}
\newcommand{\by}{{\bf y}}
\newcommand{\bC}{{\bf C}}
\newcommand{\bM}{{\bf M}}
\newcommand{\bR}{{\bf R}}
\newcommand{\bZ}{{\bf H}}
\newcommand{\bh}{{\bf h}}
\newcommand{\Det}{{\rm Det}\,}
\newcommand{\blambda}{{\mbox{\boldmath $\lambda$}}}
\newcommand{\bmu}{{\mbox{\boldmath $\mu$}}}
\newcommand{\bsigma}{{\mbox{\boldmath $\sigma$}}}
\newcommand{\brho}{{\mbox{\boldmath $\rho$}}}
\newcommand{\bxi}{{\mbox{\boldmath $\xi$}}}
\newcommand{\bdot}{{\mbox{\boldmath $\cdot$}}}
\newcommand{\bdots}{{\mbox{\boldmath $:$}}}
\newcommand{\re}{\text{\rm Re }}
\newcommand{\smb}{\mathscr{B}}

\def \I{\mathrm{i}}
\def \Ident{\mathrm{I}}
\def \Ik{I_k}
\def \Ikk{I_{k+1}}
\def \Ipar{I}
\def \J{J}
\def \R{\mathbb{R}}
\def \C{\mathbb{R}}
\def \Z{\mathbb{Z}}

\def \x{\mathrm x}
\def \A{\mathrm{A}}
\def \B{\mathrm{B}}
\def \CC{\mathrm{C}}
\def \E{\mathrm{E}}
\def \K{\mathrm{K}}
\def \L{\mathrm{L}}
\def \N{\mathrm{N}}
\def \P{\mathrm{P}}
\def \Q{\mathrm{Q}}
\def \T{\mathrm{T}}

\def \G{\mathcal{G}}
\def \LL{\mathcal{L}}
\def \PP{\mathcal{P}}
\def \QQ{\mathcal{Q}}
\def \sint{\mbox{\small$\int$}}
\def \Cov{\mathrm{Cov}}
\def \det{\mathrm{det}}
\def \diag{\mathrm{diag}}

\setlength{\textwidth}{6.25in}
\setlength{\textheight}{8.5in}
\setlength{\oddsidemargin}{.5in}
\setlength{\topmargin}{-.25in}

\title{Ensemble Dynamics and Bred Vectors}
\author{Nusret Balci${\,}^1$, Anna L. Mazzucato${\,}^2$,
Juan M. Restrepo${\,}^{3}$\footnote{{\it Corresponding Author:} 
{\tt restrepo@math.arizona.edu}},
George R. Sell${\,}^4$\\
${\,}^1$ {\em Institute for Mathematics and its Applications} \\
{\em University of Minnesota, Minneapolis MN 55455 U.S.A.} \\
${\,}^2$ {\em Department of Mathematics} \\
{\em The Pennsylvania State University, University Park, PA 16802 U.S.A.} \\
${\,}^3$ {\em Department of Mathematics and Department of Physics} \\
{\em University of Arizona Tucson, AZ 85721 U.S.A.} \\
${\,}^4$ {\em School of Mathematics} \\
{\em University of Minnesota, Minneapolis MN 55455 U.S.A.}}
\date{\today}
\maketitle
\thispagestyle{empty}

\begin{abstract}
We introduce the new concept of an EBV to assess the
sensitivity of  model outputs to changes in initial conditions for
weather forecasting. 
The new algorithm, which we call the {\em Ensemble Bred Vector} or
EBV, is based on collective dynamics in essential ways. 
As such, it keeps important geometric features which are lost in the earlier bred vector algorithm (BV).
By construction, the EBV algorithm produces  one or more dominant vectors, and is  less
prone to spurious results than the BV algorithm. It retains 
the attractive features of the BV with regard to being able to handle legacy codes,
with  minimal additional coding. 

We investigate the performance of  EBV, comparing it  to  the BV algorithm
as well as the finite-time Lyapunov Vectors. With the help of a continuous-time
adaptation of these algorithms, we  give a theoretical justification
to the observed fact that the vectors produced by BV, EBV, and the 
finite-time Lyapunov vectors are similar for small amplitudes. The continuum theory 
is establishes the relationship
between the two algorithms and general directional derivatives. 

Numerical comparisons of BV and EBV  for the 3-equation Lorenz model and for
 a forced, dissipative
 partial differential equation of Cahn-Hilliard type that arises in
 modeling the thermohaline circulation,
  demonstrate that the EBV yields a size-ordered description of the 
 perturbation field, and is more robust than the BV in the higher nonlinear regime. The 
 EBV yields insight into the fractal structure of the Lorenz attractor, and of the inertial manifold
 for the Cahn-Hilliard-type partial differential equation.
\end{abstract}

\noindent {\em Keywords:
Bred vectors, Lyapunov vectors, sensitivity, dynamic stability, Cahn-Hilliard, Lorenz.} \\
\noindent PACS 94.05sx, 92.60.Ry, 92.60.Wc, 93.65.+e, 92.70.-j, 02.50.-r

\thispagestyle{empty}

\section{Introduction}

Central to weather prediction is the analysis of the sensitivity of a physical or
computer-coded model to  initial conditions.  Model sensitivity to parameters 
is also important in  model inter-comparison. One studies such sensitivity in 
order to obtain a  better understanding of the role played by these parameters in model outcomes.

Sensitivity and predictability are often intertwined in the context
of weather prediction and have been the subject of extensive research
(see \cite{btmp} and references contained therein.) These are 
not exclusively weather-related issues and thus geophysical fluid dynamics
will often mine other physical, computational and mathematical disciplines,
for ideas with which to assess dynamic sensitivity. Practical sensitivity methodologies must
contend with the evolution and dynamics  of highly coupled, complex, high-dimensional
systems, riddled with subscale parameterizations and empirical relations, which are the norm
in large-scale climate and meteorology models.

A tool used in the study of sensitivity analysis is the
Bred Vector (BV) algorithm. It is proposed for use
in forward sensitivity of weather and climate models.
While in Subsection 2.1, we present a brief survey
of some of the applications of this algorithm in various sensitivity
analyses, in this article, we will focus on the issue of the maximal
growth of errors due to small changes in the initial conditions.

The concept of the BV algorithm we use is based on the theory
first introduced in Toth and Kalnay (1993).
In addition to the BV notion, we present here a new variant, 
which we call the {\it Ensemble Bred Vector}  (EBV) algorithm.  
The definitions of both the BV and EBV algorithms are presented in Section \ref{problem}.

In the BV algorithm, one follows an initial condition of the time-discrete
nonlinear system, along with cloud, which describes a family of nearby solutions. 
(Since this algorithm is used to sample the error space, an ensemble of initial perturbations
is bred simultaneously.) The perturbations at the initial time are fixed with
a common small amplitude $\epsilon$. After each cycle, the outcome of the perturbations
 is rescaled to the same amplitude $\epsilon$. For the BV algorithm, the
rescaling of each perturbation is independent of the others, and there is
no mechanism to use the rescaling to compare the dynamics of nearby perturbations.

The new variation that we propose here, the EBV algorithm, differs from the
BV algorithm,  in the rescaling rule. In particular, for the EBV those
perturbations that are not the same size as the largest perturbation, play a
reduced role after the rescaling. Thus the rescaling used in the EBV algorithm
 serves us better in separating various levels of the dominant dynamics. 
In short, the EBV algorithm offers better insight into the
relative behavior of nearby trajectories.
Therefore, even when the initial perturbations of the two algorithms are based on
the same cloud, the EBV algorithm is  linked to the ensemble dynamics of the underlying non-linear model,
hence its name.  We will make use of both the BV and the EBV algorithms.
In fact, one of our major goals is to present an in-depth comparison
of the two algorithms, as they are used, or can be used, in describing
the underlying dynamics of the model.

We will show that in some metrics, the BV and EBV algorithms are
comparable, with the EBV being more accurate and faster, see
Table \ref{tab1} and Figures \ref{cycase2}, \ref{compert}, \ref{perts}, for example. 
For other issues, especially those
involving the longtime dynamics within the global attractor, the BV
algorithm has a shortcoming, which limits its use (see
Section 4). 
The EBV, instead, leads to  useful and interesting insight into the
dynamics of the model, as is shown in the three Figures \ref{difflorenz2b}, \ref{zoom1}, and \ref{zooma}.

This brings up the question: For a given model of sensitivity with respect
to initial conditions, how does one determine the direction vector that
results in the maximal increase in the error due to a small perturbation
in the given direction initially? This is where the Lyapunov vectors enter the scene.
What one needs is a {\bfseries red vector}, which is the Lyapunov
vector $x_0$ with $\|x_0\|=1$ and with the property that the
corresponding strong Lyapunov exponent $\lambda_1$ is the maximal
Lyapunov exponent for the model.
(See Subsection \ref{red} for the definition and more details.
One should note that the Lyapunov exponents for the model
require integration over $0\leq t<\infty$, or over the
real line $\mathbb{R}$.)

We will use either the EBV or the BV algorithm
to approximate the solutions of the tangent linear equation.
In Section 3 we show that either algorithm is a good
approximation. For these algorithms one can
 integrate only over a finite interval $0\leq t\leq T$, where $0<T<\infty$. However,
it is only under  exceptional circumstances, {\it e.g.}, time-periodic or autonomous problems, 
 that a finite-time integration will approximate well an
infinite time average.  For any hope for
success in using a finite-time integration, we require that the model problem
satisfy two properties:
$$
\begin{aligned}
(1)\quad&\text{There is an attractor $\mathfrak{A}$ for the solutions}, \\
(2)\quad&\text{The initial condition is chosen to be on, or near, $\mathfrak{A}$}.
\end{aligned}
$$

What can one expect with such a finite-time approximation,
when the initial condtions are {\it near \,} the attractor 
$\mathfrak{A}$?
In terms of the calculated time, one expects first to be in a
transient state.
Then after a while, one hopes to get some meaningful
information about the longtime dynamics of the model.
We include in this manuscript several studies of such approximations.

It is very important to note that, in
order to better understand the maximal growth of errors
due to small changes in the initial conditions, one needs to
exploit the dynamical information contained in the attractor
of the model. In particular, one needs to complete two steps:

\begin{itemize}
\item Step 1:
One needs to locate the red vector $x_0$ and the associate
	Lyapunov exponent $\lambda_1$.
 	This then determines the red vector solution
	$R(x_0,t)$, for $-\infty<t< \infty$,
	for the model equation.	
\item Step 2: One must find a good approximation of the time
	evolution $R(x_0,t)$ on an appropriate finite-time interval,
	$0\leq t\leq T<\infty$.
\end{itemize}

Once the red vector $x_0$ is known, then $\epsilon\,x_0$
is the initial condition for the bred vector sequence,
see Subsection 2.2.
Due to the results derived here in Section 3, either $EBV(t)$,
or $BV(t)$, is a good finite-time approximation of $R(x_0,t)$,
for $0\leq t\leq T<\infty$.
This takes care of Step 2.
Consequently, the problem boils down to a search for the red vector,
which is addressed below in Subsection \ref{red}.
Fortunately for us, there is wealth of related mathematical information in
the 1987 manuscript of R. A. Johnson, K. J. Palmer, and G. R. Sell.
see \cite{JPSell87}.
(We will refer to this paper as the ``JPS87'' in the sequel.)
As we shall show, by
using the JPS87, we are able to describe the mathematical
process of finding the red vector. By using this citation, with the theory of the $EBV(t)$ algorithm,
this leads to a good solution for our sensitivity problem.

It was observed by Toth \& Kalnay {\em op. cit.}  (see also
\cite{tothkalnay97}) in 
several experimental runs that BVs resemble the {\em leading}   finite-time Lyapunov vectors. 
In  order to make a more quantitative comparison between Lyapunov
vectors and BVs, in Section 
\ref{interpret}  we
study  the Continuum Limits (as the basic time-step size for rescaling goes to 0) of the BV and
the EBV algorithms and show direct connections between these limits
and specific  solutions of the continuous-time tangent linear equations. 
Section \ref{interpret} also contains  a further discussion of  some of  the
desirable and interesting features of the new EBV. For instance,
there is a natural ordering of the ensemble members of an  EBV, and as we will show, it is
possible to observe  perturbations with smaller sizes than the dominant
one, but with very strong growth, see the spear-like behavior in Figure \ref{difflorenz2b}. 

We consider two models to exemplify the features of the new EBV. The first is the familiar
Lorenz63 model introduced by \cite{lorenz63}. It has a well-known global attractor. 
 The second is a nonlinear forced and dissipative partial differential equation 
 of the  Cahn-Hilliard type. This equation is a variation of a model proposed  by
\cite{cy92} of the oceanic thermohaline circulation. We will denote
the equation associated with this  model as  the CY92 in this
study. (In fact, we impose periodic boundary conditions instead of the more physical zero-flux, zero-stress
conditions at the poles.)

It turns out, it is a good example of the typical
climate-related model dynamics. However, the CY92 is special, since it has an inertial
manifold.  Consequently, the longtime dynamics of this partial differential equation
is completely contained in the attractor of a finite dimensional
ordinary differential equation.  
In applications, the BV algorithm and its variants are in fact discrete-time algorithms based
on finite-dimensional  approximations of weather models, obtained
either by mode projection as in the \cite{lorenz63} system, or by
spatial discretization of  partial differential equations (PDEs), as in the Cessi-Young (CY92) model, which 
will be described below.
As we will show the EBV yields insights into the 
structure of the  attractor of the Lorenz63, and of the inertial manifold to the 
CY92.

Our numerical examples also highlight that the BV algorithm is sensitive to the amplitude
and frequency content of the initial perturbation.
 In contrast, the outcome  of the EBV algorithm shows a clear hierarchy among
its members, and the first few  members already
generate an unambiguous  characterization  of the
perturbation field  at both large and small amplitudes. In the  nonlinear 
regime, however, the EBV will be shown in Section \ref{examples} to be less likely to 
produce spurious results than the BV. In Section \ref{conc}  we will address 
implementation issues of the EBV.

\section{BV and EBV Algorithms; Finite-Time Lyapunov Vectors}
\label{problem}

\sep
In this section we present the definitions and methodology
for computing the two Bred Vector algorithms, the BV and the EBV. We also
review the basic theory of Lyapunov Vectors,  Lyapunov exponents, and their finite-time counterparts.
We will compare these different tools below. However, before doing this, we include here a brief
survey of some of the applications and theoretical issues that have been
noted in the implementation of the BV algorithm.

\subsection{Brief Survey of Applications of the BV Algorithm}

For the BV algorithm, we use the one originally proposed by
\cite{tothkalnay93}. 
BV  is purely algorithmic. 
It is ``equation-free'' and thus with additional minimal computer coding it can handle legacy code representing even extremely complex models. 
Most alternatives for obtaining estimates of forward sensitivity 
will involve non-trivial additional coding. 
For example, in order to obtain the finite-time  Lyapunov vectors and exponents,
one needs to derive and make use of the tangent linear model. (From the beginning, it was realized that 
there existed some close connections
between the BV algorithm and the tangent linear equations. As we will show in
Section 3, there is a rigorous mathematical foundation for these connections.)
Singular value decomposition methods, which can offer complementary information to 
Lyapunov-vector inspired methods, also require a tangent linear model.
 \cite{ddamg}  use this approach to study regime
predictability in some reduced weather models.
Other examples are: \cite{btmp} and \cite{pgbb}. 
In another direction, \cite{WolfeSamelson07} propose the use 
of the MET  and the finite-time singular vectors to approximate the Lyapunov vectors.

The BV algorithm  is a finite-time, forward sensitivity methodology which, in addition to 
being useful in  characterizing
model sensitivity to initial conditions, has been proposed as a means to produce a 
reduced-rank representation of the background error 
in data assimilation and forecast error-covariance approximations (see \cite{corazza03}, for example).

Several articles in the  literature have addressed applications of  the BV algorithm in weather modeling. See, in particular \cite{tothkalnay93}, \cite{tothkalnay97}, \cite{kalnaybook}, and reference therein. 
For a comparison of  the BV algorithm and other methods, such as Monte-Carlo perturbed observations, we refer for example to \cite{Cheung01,GneitRaft05,HansenSmith00,WeiToth03}.
For an application of the BV algorithm to   ensemble Kalman filters see \cite{WangBishop03}.
\cite {PrimoEtAl08} and \cite{Hallerberg} have treated applications based on
variations in the algorithm, where for example, the rescaling is done by using a
geometric mean.

As already discussed, several alternatives to the BV algorithm 
have been proposed and employed in the literature.  BVs have been viewed as non-linear analogs of finite-time Lyapunov vectors. 
Similarly nonlinear analogs of  singular vectors have been proposed, for instance conditional nonlinear optimal perturbations proposed by \cite{MuJiang08} and non-linear singular vectors proposed by \cite{RiviereEtAl08}, although they 
 entail a computationally expensive optimization.

Both the BV and EBV algorithms arise in the time-discretization of a
continuous-time dynamical system.  We consider the following initial value problem
\begin{align} \label{EQ1}
\begin{aligned}
\frac{dy}{dt} &= G(y), \quad t >0,
\\
y(0) &= y_0,
\end{aligned}
\end{align}
where $t$ represent time and $G = G(y) $ is a map that has at least a bounded gradient. Since our main
applications involve autonomous differential equations,
we assume that $G$ does not explicitly depend on time. 
The basic theory we present here has a routine extension to
non-autonomous problems.

The solution vector $y=y(t)$ can live in a finite- or infinite-dimensional normed linear space. 
In the former case, \eqref{EQ1} is an (autonomous) system of ordinary differential equations, 
while in the latter case, \eqref{EQ1} is an (autonomous) system of partial differential equations, 
modeling a time dependent, spatially extended system.
For systems of partial differential equations, we assume that
either periodic boundary conditions or non-flux boundary conditions
 are prescribed. See for example, Sell and You (2002). The use of other boundary conditions 
may lead to a related theory, but we do not
address the issue here.

If the system contains evolution partial differential equations, the system, along with the boundary conditions, are  discretized in space or projected 
onto a finite-dimensional space compatible with the boundary conditions.
 Consequently,  we usually  assume that \eqref{EQ1} is a system of ordinary differential equations of dimension $K$, which may be large.  
Since most  large-scale weather and climate circulation models  presently use explicit-in-time integrators, we will focus on numerical models of this type.

\subsection{Bred Vector Algorithms: BV and EBV} \label{sec.BV}


We let  $y=y(t)$ denote a given 
(continuous time) solution of 
\eqref{EQ1}. We then turn to an approximate solution
$Y_n\,=\,Y(t_n)$, which is defined on the time grid:
$t_n$, for $n=0,1,2,\cdots$, where 
$t_{n+1}=t_n+\delta t_n$, for $n=0,1,2,\ldots$. We set $t_0=0$.
We assume that $\delta t_n=\delta t$ is positive and small,
and that it does not depend on $n$, for $n\geq 0$.

For the autonomous case, the initial value problem, which is 
  approximated using an explicit numerical integration scheme, leads
to consideration of the difference equation
\begin{align} \label{EQ2}
\begin{aligned}
Y_{n+1} &= M(Y_n,\delta t), \quad n=0,1,2,..,\\
Y_0 &= Y(0).
\end{aligned}
\end{align}
where $Y(t_n)$ is a solution of the discrete problem \eqref{EQ2}, and it may be viewed
as an approximation of the continuous-time solution $y(t)$, at $t=t_n$.
Likewise, the initial condition  $Y_0$ is an approximation of the initial condition  $y_0$.
The points $Y_n$ are in the 
$K$-dimensional Euclidean space $\mathbb{R}^K$ and
$M=M(Y,\delta t)$ is the discrete-time solution operator on \ $ \mathbb{R}^K$ generated by the ordinary differential equation 
(\ref{EQ1}).
As noted above, we assume that the discrete-time problem \eqref{EQ2} has 
an attractor $\mathfrak{A}$, and that $Y_0$ is on, or near $\mathfrak{A}$.

At this point it is convenient to introduce the related concepts of a
{\it Cloud} (at $t=0$) and a {\it Family of Initial Perturbations}.
A Cloud  (at $t=0$) is a family of tangent vectors 
$\delta\mathcal{Y}_0(\iota)$ to $Y_0$  in $\mathbb{R}^K$ that depends on $\iota$,
where $\iota\in\mathbb{I}$ and $\mathbb{I}$ is a finite index set.
The main requirement we impose is that
$$
\|\delta\mathcal{Y}_0(\iota)\| \,=\,\epsilon,
\qquad\text{for all }\iota\in\mathbb{I},
$$
where $\epsilon$ is small, positive, and fixed.
The collection of all terms $(Y_0,\delta\mathcal{Y}_0(\iota))$ 
in $\{Y_0\}\times \mathbb{R}^K$, for $\iota\in\mathbb{I}$,
is called a Family of Initial Perturbations.
Notice that this collection lies on a sphere of radius $\epsilon$
in $\{Y_0\}\times\mathbb{R}^K$ with center at $(Y_0,0)$. Both the Cloud and the
perturbations evolve in time, via the BV or the EBV algorithms,
which are defined below.
As we now note, the definitions of the two algorithms differ only in
the rescaling rule.

We begin by recalling the BV algorithm as given by Toth\&Kalnay.
For $n \geq 0$, we assume that the base point $Y_n$ and the
perturbation vector $\delta Y_n(\iota)$ are known. For the $(n + 1)^{\text{st}}$ step we use:
\begin{enumerate}
\item $Y_{n+1}$ denotes the
$(n+1)^{\text{st}}$ base point, and it
is determined by \eqref{EQ2};
\item $\delta\mathcal{Y}_{n+1}(\iota)$,
the $(n+1)^{\text{st}}$ perturbation vector, is given by
\begin{align}
\label{EQ3}
\delta Y_{n+1}(\iota)\,&=\,M(Y_n + \delta\mathcal{Y}_{n}(\iota),\delta t) - M(Y_{n},\delta t),\\
\delta\mathcal{Y}_{n+1}(\iota)\,&=\,R_{n+1}\,\,\delta Y_{n+1}(\iota),
\end{align} 
where $R_{n+1}$ is a rescaling rule. 
\end{enumerate}
The time evolution of the Cloud is BV$(t_n) = \delta {\cal Y}_n(\iota)$, where $\iota \in \mathbb{I}$ and $n \geq 0$.

One rescaling rule, proposed by 
Toth and Kalnay, {\em op. cit.}, consists of rescaling the
$n^{\text{th}}$  perturbation vector  to the previous one by
\begin{equation}
\label{BVeq}
\|\delta \mathcal{Y}_{n+1}(\iota) \|= R_{n+1} \|\delta Y_{n+1}(\iota)\|
	=\|\delta \mathcal{Y}_0(\iota)\| =\epsilon,
\quad\text{for }\iota\in\mathbb{I}, n\geq0.
\end{equation}  
Equivalently, 
\begin{equation}
\label{rule1}
R_{n+1}(Y_0) :=
	\frac{\|\delta\mathcal{Y}_n(\iota)\|}{\|\delta Y_{n+1}(\iota)\|}
	\,=\,\frac{\|\delta\mathcal{Y}_0(\iota)\|}
		{\|\delta Y_{n+1}(\iota)\|},
\end{equation}
so that   $R_{n+1}=R_{n+1}(Y_0)$ 
depends on the initial base point $Y_0$, as well as the
initial perturbation vector
$\delta\mathcal{Y}_0(\iota)$.
An alternate rescaling rule consists in rescaling periodically,  
at $t_{m\,k}$, where $k$ is an integer $k\geq2$,
and $m=1,2,\cdots$.
In this case, one uses the rule \eqref{rule1} when
$n=mk$, and
\begin{equation}
\label{rule2}
R_{n+1}(Y_0)\,=\,1,
\qquad\text{for }m k< n<(m+1)k.
\end{equation}
We do not use this alternate rule in this paper.
(For a discussion of rescaling time and regime predictability  in some reduced models, 
see \cite{ddamg} and references therein).

For the {\it Ensemble Bred Vector} algorithm,  instead of using
$R_{n+1}$ as in equation \eqref{EQ3},
we use a uniform scaling
$\boldsymbol{\large {R}_{n+1}^{\text{min}}}$, 
which is the same for
all $\iota\in\BI$.
In particular, we replace equation \eqref{EQ3} with
\begin{equation}
\label{ebv}
\begin{gathered}
\delta Y_{n+1}(\iota)\,
	=\,M(Y_n+\delta\mathcal{Y}_n(\iota),\delta t)-M(Y_n,\delta t) \\
\delta\mathcal{Y}_{n+1}(\iota)\,=\,
\boldsymbol{\large {R}_{n+1}^{\text{min}}}\,
\delta Y_{n+1}(\iota),
\end{gathered}
\end{equation}
for all $\iota\in\BI$, where
\begin{equation}
\label{r-bold}
{\bf {\large R}}_{n+1}^{\text{min}}\,=\,
\epsilon \left[\max_{\iota\in\BI}(\|\delta Y_{n+1}(\iota)\|)
\right]^{-1}.
\end{equation}

Similarly, when \eqref{ebv}
and \eqref{r-bold} hold, we use EBV$(t_n) = \delta {\cal Y}_n(\iota)$, for
$\iota \in  \mathbb{I}$ and $n \geq  0$, to denote the time evolution of the Cloud.
(Alternatively, a periodic rescaling rule utilizing \eqref{rule2} above can be employed.) 
The time step used to compute the base trajectory and the time intervals between normalizations need
not be the same. This is the case for both the BV algorithm as well as EBV.

The crucial difference between the BV and EBV algorithms is that, 
even when the BV is run concurrently over an ensemble of initial data, 
the outcome of the algorithm for each given datum does not depend on 
the other members of the ensemble. In contrast, the evolution of the ensemble members is
interdependent in the EBV. Nevertheless,
 by construction,  the EBV should exhibit similar 
behavior to the BV at small amplitudes. Indeed, one of the design principles for
the EBV algorithm is to reinforce this aspect by providing it with a built-in acceleration mechanism.

Both the BV algorithm  and EBV outcomes, on the other hand 
 depend on the choice of vector norm used to define the rescaling rule
 in either \eqref{rule1} or \eqref{r-bold}. 
While all norms are equivalent in the (finite) $K$-dimensional space
where we seek solutions, in practice the constants appearing in the
equivalence between different finite-dimensional norms generally
strongly depend on the dimension $K$ and eventually blow up as $k$
becomes infinite.  
Hence, the choice of norm used can have an impact on the implementability and performance of these algorithms.
In  \cite{RiviereEtAl08}, it was 
suggested   that the outcome of the BV algorithm may depend strongly  on the choice of norm.
This is actually a consequence of the non-negligible nonlinear effects in the system.
In Section \ref{conc}, we will elaborate  further on the issue of norm dependence.

\sep

In addition to the ``renormalization time-step'' $\delta t$,
there is another time step, the ``integration time-step'',
which we will denote by $\Delta t$.
For example, one encounters the new time step when moving
from the continuous-time problem equation (1) to the discrete-time
problem equation (2).
For the most part, we will treat $\Delta t$ and $\delta t$ as being
equal in the calculations described in this article. 
However, we always require that $\Delta t \leq\delta t$.

\subsection{Lyapunov Vectors} \label{sec.LV}

In Section \ref{examples}, we will be applying  the BV and EBV algorithms
to two models 
consisting of systems of (nonlinear) autonomous ODEs of the form \eqref{EQ1}. 
( The second model arises from the discretization of a PDE.)
 In each model, there is a compact, global  attractor $\fma$, which is
 a subset of $\R^K$, and is  invariant for the time evolution of the system.
(We refer the reader to 
Chapter 2 in \cite{SY02} for more information on attractors and
global attractors.)
We will let $\theta$ denote a typical point in the attractor $\fma$,
and we will let $\theta\cdot t\,=\,y(t)$ denote the unique solution
of \eqref{EQ1} that satisfies $y(0)=\theta$.
Since $\fma$ is invariant, one has $\theta\cdot t\in\fma$, for 
all $t\in \BR$.

\sep
In order to study the sensitivity with respect to initial
conditions on the attractor, the Tangent Linear Model is used, which
is defined as 
\begin{subequations}
\label{ode1}
\begin{equation} \label{ode1.a}
   \partial_t x\,=\,A(\theta\cdot t)\,x, 
\end{equation}
\begin{equation} \label{ode1.b}
  x(0)\,=\,x_0\in\BR^K,
\end{equation}
\end{subequations}
 where $A(y)\,=\,DG(y)$ is the Jacobian matrix of $G$.
Hence $A\,=\,A(\theta\cdot t)$ is the linearization of \eqref{EQ1}
along the solution $\theta\cdot t$. 
We observe that, even if $G$ does not explicitly depend on time, \eqref{ode1.a} is generally non-autonomous,
since $\theta\, \cdot\, t$ changes with time.

We let $U(\theta,t)$, denote the solution operator of \eqref{ode1.a}, which takes the initial data  to the solution at time $t$,  so that 
$U(\theta,t)x_0$ is the solution of  the initial value problem for \eqref{ode1}. \ Such an operator is well defined by the uniqueness of solutions to the problem \eqref{ode1}. 
Uniqueness of solutions also readily implies the cocycle identity:
\begin{equation}
\label{cocy}
U(\theta,\tau+t)\,=\,U(\theta\cdot \tau,t)\,U(\theta,\tau),
\qquad\text{for all }\theta\in\fma\text{ and all }\tau,\,t\in\BR.
\end{equation}

\sep
Next we consider a family of mappings $\Pi=\Pi(t)$, which are defined
for $t\in\BR$ by the relation
\begin{equation}
\label{Pi}
\Pi(t)(\theta,x_0)\, := \,(\theta\cdot t,U(\theta,t)x_0),
\qquad\text{for }(\theta,x_0)\in\fma\times \BR^K.
\end{equation}
We note that $\Pi(t)$ maps $\fma\times\BR^K$
into itself, for each $t\in\BR$;
it is jointly continuous in $(t,\theta,x_0)$; 
it satisfies $\Pi(0)(\theta,x_0)=(\theta,x_0)$, 
(i.e., $\Pi(0)=I$, the identity operator;
as well as the evolution property:
\begin{equation}
\label{group}
\Pi(\tau+t)\,=\,\Pi(t)\,\Pi(\tau),
\qquad\text{for all }\tau,\,t\in\BR.
\end{equation}

By using the discrete-time dynamics, where $t$ and $\tau$ are restricted to
satisfy $t=n\cdot\delta t$ and $\tau=m\cdot\delta t$,
the notation and the theory of dynamical 
systems extends readily to the discrete-time problems of interest herein.
Note that the $\theta$-component of $\Pi$ does {\it not} depend on
the $x_0$-component. Thus $\Pi$ is called a 
{\em skew product} flow. Since $\Pi$ is linear in $x_0$,
it is sometimes called a {\em linear} skew product flow.
In summary, the Tangent Linear Equation over the attractor $\fma$
generates a linear skew product flow. (For more information
on the theory of skew product flows in the context of
non-autonomous dynamics, see the multiple works of
Sacker and Sell, for example: \cite{SS77}, \cite{SS78} and \cite{SS80}.)
The dynamics of $\Pi$ are crucial for understanding
the sensitivity and predictability of the underlying model.

In his opus magnum, which was published in Ukraine in
1892, Lyapunov presented his theory of stability for 
finite-dimensional ordinary differential equations.
This work includes his study of the non-autonomous linear problem \eqref{ode1},
see \cite{AML92} (yes, $100$ years later.)
One of Lyapunov's goals was to develop an analogue of
the well-known eigenvalue-eigenvector theory, for the 
solutions of the autonomous problem, to the study of solutions
of general non-autonomous equation \eqref{ode1}.

The approach developed by Lyapunov begins with the 4 Lyapunov Relations
of exponential growth:
$$
\limsup_{t\to\pm\infty}\frac1{t}\log(\|U(\theta,t)x_0\|) 
\quad\text{and}\quad
\liminf_{t\to\pm\infty}\frac1{t}\log(\|U(\theta,t)x_0\|),
$$
where $x_0\not=0$.
Lyapunov was interested in, as are we, the case where these four limits 
are equal, and
\begin{equation}
\label{strong}
\lambda(\theta,x_0)\,\eqdef\,
\lim_{t\to -\infty}\frac1{t}\log(\|U(\theta,t)x_0\|)\,=\,
 \lim_{t\to \infty}\frac1{t}\log(\|U(\theta,t)x_0\|),
\end{equation}
where $x_0\not=0$.
The linearity of $U(\theta,t)x_0$  implies that
$s\,\lambda(\theta,x_0)=\lambda(\theta,s\, x_0)$, for $s\not=0$.
Hence one can assume, as we do, that $x_0$ is a unit vector,
i.e., $\|x_0\|=1$.
When \eqref{strong} holds, then $\lambda(\theta,x_0)$ is a {\it strong
Lyapunov exponent}, and the unit vector $x_0$ is an (associate)  {\it Lyapunov vector}.
The {\it Lyapunov spectrum}, LY$\Sigma$,
is the collection of all such $\lambda(\theta,x_0)$,
with $\theta\in\mathfrak{A}$ and $\|x_0\|=1$. 
For example, if $A(\theta\cdot t)=A_0$ is an autonomous matrix, 
then the Lyapunov spectrum consists of all real numbers $\lambda$ that satisfy
$\lambda\,=\,\text{\rm Re}\,\nu$, where $\nu$ is an eigenvalue of $A_0$.

A Lyapunov vector is not an isolated vector, rather it spawns a line
of Lyapunov vectors (through the origin) in $\R^K$.
That is to say, a Lyapunov vector is a {\em point} in $\mathcal{P}^{K-1}$,
the $(K-1)$-dimensional projective space.
For a given vector $v\not=0$ in $\R^K$, we will use 
$[v]\in\mathcal{P}^{K-1}$
to denote the unique line in $\R^K$
that contains $v$. (Note that $[v]=[-v]$.)
Conversely, when one maps a line $[u]$ in $\mathcal{P}^{K-1}$
to a vector $u\in \R^K$, we require that the pre-image $u$ lie on
the line $[u]$ and that $\|u\|=1$.
One should note that $\mathcal{P}^{K-1}$ is a metric space.
The {\it projective metric} $d_{\rm pr}(u_1,u_2)$ is defined for
nonzero vectors $u^1$ and $u^2$ in $\mathbb{R}^K$ by
\begin{equation}
\label{d-pr}
d_{\rm pr}(u^1,u^2) \, = \, \min_{s_1,s_2} \|s_1 u^1 \pm s_2 u^2\|,
\end{equation}
where $s_1$ and $s_2$ are real numbers that satisfy
$\|s_1u^1\| = \|s_2u^2\| = 1$.
This metric is used for measuring the distance between the
lines $[u^1]$ and $[u^2]$ in $\mathcal{P}^{K-1}$.

\sep
Since the
solution operator $U(\theta, t)$ of the linear problem \eqref{ode1}
maps lines in $\R^K$ onto lines, one can use this operator
to define a related {\it projective flow} $\Sigma(\theta, t)$ on $\mathcal{P}^{K-1}$
by means of the relation
$$
\Sigma(\theta, t)\,[u]\,=\,[U(\theta, t)\,u],
\qquad\text{for }[u]\in\mathcal{P}^{K-1}.
$$
Using this, one obtains an equivalent flow $[\Pi]$
on $\mathcal{P}^{K-1}\times\fma$, where
\begin{equation}
\label{[Pi]}
[\Pi](t)(\theta,[x_0])\,=\,(\theta\cdot t,\Sigma(\theta, t)[x_0]),
\qquad\text{for }(\theta,[x_0])\in\fma\times\mathcal{P}^{K-1},
\end{equation}
compare with \eqref{Pi}.
One obtains additional information about the dynamics
on the projective flow $[\Pi(t)]$, by using the Lyapunov vectors, as is
noted below.

\subsection{The Finite-Time Lyapunov Vectors}

Next we turn our attention to the question of finding 
good finite-time approximations of these Lyapunov vectors.

 We begin by constructing a piecewise autonomous approximation of the Tangent Linear Equation for \eqref{EQ1}. To this end, we replace \eqref{ode1} by
\begin{subequations}
\label{EQ5}
\begin{equation}
\label{EQ5.a}
\partial _t  Y(t)\,=\, A_n\,Y(t), \text{ for } t_n < t \le t_{n+1}, 
\end{equation}
\begin{equation}
\label{EQ5.b}
Y(0) = Y_0,
\end{equation}
\begin{equation}
\label{EQ5.c}
t_n = n \, \delta t, \qquad n=0,1,2,\cdots,N-1.
\end{equation}
\end{subequations}
where $A_n=A(y(t_n)) : = \frac{\partial G}{\partial y}|_{y=(y(t_n))}$ 
and $y(t_n)$ is the value of an exact solution of \eqref{EQ1} at $t=t_n$. The solution of this system at the grid points $t_n$ is, explicitly and recursively, given by
\begin{equation}
\label{EQ6}
Y_{n+1}\,=\,e^{\delta t\, A_n}\,Y_n,
\qquad\text{for }n=0,1,2,\cdots,N-1.
\end{equation}
Consequently, $Y_N$, the solution at time $T=N \, \delta t$,
is given by
\begin{equation}
\label{sol-lp}
Y(T)\,=\,Y_N\,=\,Z(T)\,Y_0,
\end{equation}
where
$$
Z(T)\,:= 
\,e^{\delta t\,A_{N-1}}e^{\delta t\,A_{N-2}}\cdots e^{\delta t\,A_{1}} e^{\delta t\,A_{0}}.
$$
We now define the 
 approximation of the Lyapunov Vector associated with
the largest Lyapunov Exponent $\lambda_1$, 
at the time $t=T$ - {\em the finite-time Lyapunov vector},
as the direction of steepest ascent for the matrix $Z(T)$. 
For example, if one had used an explicit Euler scheme, then $W(T)$ would be an Eulerian
approximation of $Z(T)$, where
$$
W(T)=(I+\delta t\,A_{N-1})\cdots (I+\delta t\,A_1)\cdots
	(I+\delta t A_0).
$$
The finite-time Lyapunov Vector is the singular vector corresponding to the largest singular value
of  $Z(T)$.

In practice (see e.g. Section \ref{cy92} for the case of the CY92 model), the finite-time LV will be computed by directly solving a discrete approximation of the LTM and rescaling the output (the rescaling can be done at arbitrary intervals of time, since the problem is linear.)

\subsection{The Search for the Red Vector} \label{red}

One of the main contributions found in the JPS87 manuscript is an
indepth study of the interactions between two major theories of the
longtime dynamics of nonautonomous, linear differential systems.
Dynamics of nonautonomous, linear differential equations:

\begin{itemize}
\item Exponential Dichotomies (and Continuous Foliations) and
\item the MET (Multiplicative Ergodic Theorem) and Ergodic Measures.
\end{itemize}

We view the JPS87 manuscript as a toolkit to be
used in the analysis of the dynamics of related  linear systems:
It is this united theory, as we now show, that forms the mathematical
foundations of the theory and applications of bred vectors.
We begin with the first aspect: Exponential Dichotomies and Continuous 
Foliations.

Consider the family of shifted semiflows
 $$
 U_\lambda(\theta,t)\,\eqdef\, e^{-\lambda t}\,U(\theta,t),
 \qquad\text{for }\lambda\in\mathbb{R},
 $$
 and the associate skew-product flows
 $$
 \Pi_\lambda(t)(\theta,x_0)\,\eqdef\,(\theta\cdot t, U_\lambda(\theta,t)\, x_0).
 $$
 Let $\mathcal{M}$ be a compact invariant set in $\mathfrak{A}$.
 As noted in \cite{SS78}, the skew-product flow $\Pi_\lambda$
 is said to have an {\it exponential dichotomy} over $\mathcal{M}$, 
 provided that there exist projectors $\mathbb{P}_\lambda$ and 
 $\mathbb{Q}_\lambda$ and constants $K_0\geq1$ and $\alpha>0$, such that
 $P_\lambda(\theta)+Q_\lambda(\theta)\,=\,I$,
 and such that, for all $\theta\in\mathcal{M}$
 and $u\in\mathbb{R}^K$, one has
 $$
\begin{aligned}\label{ed1}
\| U_\lambda(\theta,t) \, Q_\lambda(\theta) u \|   
	\leq K_0 \, e^{-\alpha t} \|u\|, \qquad t \geq 0   \\
\| U_\lambda(\theta,t) \, P_\lambda(\theta) u \|   
	\leq K_0 \, e^{\alpha t} \|u\|, \qquad t \leq 0.
 \end{aligned}
$$
When there is an exponential dichotomy and \eqref{ed1} is valid,
then the {\it stable\,} and {\it unstable\,} linear spaces,
$\mathcal{S}_\lambda$ and $\mathcal{U}_\lambda$ - which are
respectively the ranges of the projectors $Q_\lambda$
and $P_\lambda$ - satisfy important dynamical properties
that describe the exponential growth rate of selected 
solutions. For example, the system \eqref{ed1} is equivalent to:
\begin{equation}
\label{ed2}
\begin{aligned}
\| U_\lambda(\theta,t)u\| 
	&\leq K_0\,e^{-\alpha t}\|u\|, \qquad \text{for }t\geq0,\, u\in
	\mathcal{R}(Q_\lambda(\theta)), \\
\| U_\lambda(\theta,t)u\| 
	&\leq K_0\,e^{\alpha t}\|u\|, \qquad\text{for } t\leq0,\, u\in
	\mathcal{R}(P_\lambda(\theta)).
\end{aligned}
\end{equation}
Let SS$\Sigma$ denote the
SS Spectrum (aka Sacker-Sell Spectrum) for $\Pi$, which is defined as the
collection of all $\lambda \in \mathbb{R}$ such that $\Pi_\lambda$ does
not have an exponential dichotomy over $\mathcal{M}$.
 
 The Spectral Theorem in \cite{SS78} describes the continuous
 foliation, and other properties of the flow $\Pi$ over $\mathcal{M}$.
 More precisely, there is an integer $\ell$, where $1\leq\ell \leq K$,
 such that SS$\Sigma$ is the union of $\ell$ closed, bounded intervals,
 that is,
 $$
 \text{SS}\Sigma \, =\,  \bigcup_{i=0}^{\ell-1}  \, [a_{\ell-i}, b_{\ell-i}],
 \qquad\text{with }a_{\ell-i} \leq b_{\ell-i}  < a_{\ell-i-1}.
$$
Also there is a continuous foliation
$$
\mathbb{R}^K = \bigoplus_{i=0}^{\ell-1} V_{\ell-i}(\theta),
\qquad \theta\in\mathcal{M},
$$
where $\{V_{\ell}(\theta), \cdots, V_1(\theta)\}$ is a linearly independent, continuous
family of subspaces of $\mathbb{R}^K$ with 
$\sum_{i=0}^{i=\ell-1}\text{dim }V_{\ell-i}(\theta) = K$, for $\theta\in\mathcal{M}$.
As is shown below, the right-most interval $[a_1,b_1]$
plays a special role in the study of bred vectors.

For $1\leq k\leq\ell$, we let $[a_k,b_k]$ denote the $k^{th}$
spectral interval and let $V_k(\theta)$ denote the corresponding
subspace given by the continuous foliation.
We next explore the important connections between the exponential
growth rates of solutions with initial conditions in
$V_k(\theta)$ and the $k^{th}$ interval $[a_k,b_k]$.
Among other things, we will encounter the Monotonicity Property
and the Strictly Monotone Property.

Let $\nu$ and $\lambda$ be real numbers 
that satisfy $\nu<\lambda$, and both $U_\nu(\theta,t)$ and
$U_\lambda(\theta,t)$ have exponential dichotomies over $\mathcal{M}$.
Then the Monotonicity Property holds:
\begin{equation}
\label{mono1}
\mathcal{S}_\nu \subseteq \mathcal{S}_\lambda,\, \quad\text{while }\quad
\mathcal{U}_\lambda \subseteq \mathcal{U}_\nu.
\end{equation}
The Strictly Monotone Property is a consequence of the observation 
that the following three statements are equivalent:

\begin{itemize}
\item One has \quad $\mathcal{S}_\nu \not\subseteq \mathcal{S}_\lambda$.
\item One has \quad $\mathcal{U}_\lambda \not\subseteq \mathcal{U}_\nu$.
\item There is a spectral interval $[a_k,b_k]$ in the interval
	$(\nu,\lambda)$.
\end{itemize}

This brings us to a basic property.
Let $b_{\ell+1}$ and $a_0$ satisfy: $b_{\ell+1}<a_{\ell}$
and $b_1<a_0$.
Next we fix $\nu$ and $\lambda$ so that
$$
b_{k+1}<\nu<a_k\leq b_k<\lambda<a_{k-1}.
$$
Then neither $\nu$ nor $\lambda$ lie in the Sacker-Sell
spectrum $SS\Sigma$.
Furthermore, the interval $(\nu,\lambda)$ contains the spectral
interval $[a_k,b_k]$. By the Strictly Monotone Property,
the space $\mathcal{S}_\lambda$ is larger than $\mathcal{S}_\nu$,
while $\mathcal{U}_\nu$ is larger than  $\mathcal{U}_\lambda$.
Moreover, as is shown in Sacker and Sell (1978), one has:
\begin{equation}
\label{VSU}
V_k(\theta)\quad=\quad \mathcal{U}_\nu(\theta) \,
	\cap \, \mathcal{S}_\lambda(\theta),
\qquad\text{for all }\theta\in\mathcal{M}.
\end{equation}

It is a consequence of the relations \eqref{VSU} and \eqref{ed2}
that if the initial condition $(\theta,x_0)$ satisfies 
$x_0\in V_k(\theta)$, then the solution 
$U(\theta,t)x_0$ satisfies
$$
\begin{gathered}
\|U(\theta,t)x_0\| \leq K_0\,e^{(\lambda-\alpha)t}\|x_0\|,
	\qquad\text{for }t\geq 0, \\
\|U(\theta,t)x_0\| \leq K_0\,e^{(\nu-\alpha)t}\|x_0\|,
	\qquad\text{for }t\leq 0.
\end{gathered}
$$
(Note that $K_0$ depends on the choice of $\nu$
and $\lambda$.)

It should be noted that all the terms used above, including 
$\mathbb{P}_\lambda(\theta)$
and $\mathbb{Q}_\lambda(\theta)$, vary continuously in $\theta$. 
Furthermore, the exponential dichotomy is robust, in the sense that 
it varies continuously under small perturbations.
Small changes in the model result in a related exponential dichotomy with
small changes in $\mathbb{P}_\lambda$, $\mathbb{Q}_\lambda$, $K_0$,
and $\alpha$, see \cite{plsell}.

Moreover, it is shown in \cite{SS176} that if $\lambda\in$ SS$\Sigma$,
then there is a $(\theta,x_0)\in\mathcal{M}\times \mathbb{R}^K$, such that 
$\|x_0\|\ne 0$ and the solution $U_\lambda(\theta,t)x_0$ satisfies:
\begin{equation}
\label{bounded}
\sup_{t  \,\in \, \mathbb{R}} \|U_\lambda(\theta,t)x_0\| < \infty.
\end{equation}
What will become apparent shortly is that the red vector must be in the space
$V_1(\theta)$. 
Furthermore, since $\lambda\,=\,b_1$ is in SS$\Sigma$, the pair $(\theta,x_0)$, that
arises in \eqref{bounded} for this choice of $\lambda$, is a candidate for the
``red vector'' designation. A red vector must satisfy \eqref{bounded},
but the converse need not be true. More on this later.

\sep
As noted above, the second tool to be used in the theory of Lyapunov
exponents/vectors is the Multiplicative Ergodic Theorem (MET) and the ergodic measures on $\mathcal{M}$.  
One finds in JPS87 a study of the links between the LY$\Sigma$ and the SS$\Sigma$.
While the SS$\Sigma$ leads to a continuous foliation, as noted above, the MET
leads to a ``measurable'' refinement of this 
continuous splitting,
as is noted in Remark 4.2.9 on pages 177-179 in
\cite{larnold}. The latter reference is noteworthy
because it contains various extensions of the MET to problems
not originally envisioned in the pioneering works of an
earlier generation.

{\em The Multiplicative Ergodic Theorem\/}: Let $\mathcal{M}$ be a (non-empty)
compact, invariant set on the attractor  $\mathfrak{A}$,
and let $\mu$ be an ergodic measure
on $\mathcal{M}$ with $\mu(\mathcal{M})=1$. 
Then there is an invariant set $\mathcal{M}_\mu$ in $\mathcal{M}$,
with $\mu(\mathcal{M}_\mu)=1$, and there is a $k$, with 
$ \ell \leq k \leq  K $,
such that the following hold:

\begin{enumerate}
\item There is a measurable foliation 
$$
\mathbb{R}^K \, = \, \bigoplus_{j=0}^{j=k-1} \, W_{k-j}(\theta),
\qquad\text{for } \theta \in \mathcal{M}_\mu,
$$
where $\{ W_{k-j}(\theta), \cdots, W_1(\theta) \}$
is a linearly independent, measurable family of subspaces of 
$\mathbb{R}^K$ with $\text{dim }W_j(\theta)=m_j \geq 1$, 
for $1\leq j\leq k, (m_1+\cdots +m_k)=K$,
and all $\theta\in\mathcal{M}_\mu$.

\item There are real numbers $\lambda_j$,
for $1\leq j\leq k$, that are the strong Lyapunov exponents
$\lambda(\theta, x_0)=\lambda_j$,
for all $(\theta,x_0) \in \mathcal{M}_\mu \times W_j(\theta)$,
with $\|x_0\|=1$, and one has $\lambda_k < \cdots < \lambda_1$.

\item The Ergodic Spectrum LY$\Sigma(\mu)$, which depends on
the ergodic measure $\mu$, is this collection 
$\{\lambda_1,\cdots\lambda_k\}$. The Lyapunov Spectrum
$ \text{LY}\Sigma$ is the union
$$
\text{LY}\Sigma \,\eqdef \, \cup_{\mu} \text{LY}\Sigma(\mu),
$$
over all ergodic measures $\mu$, is used below.

\item For each $j$, the measurable vector bundle
$$
\mathcal{M}_\mu \times W_j(\theta)=\{(\theta,x_0)\in \mathcal{M}_\mu \times W_j(\theta) \}
$$
is an invariant set for the projective flow.
Furthermore, because of the exponential separation between these
vector bundles, the bundle $\mathcal{M}_\mu\times W_j(\theta)$, with $j=1$,
is an attractor for the projective flow.
\end{enumerate}

Connections between the SS$\Sigma$ and LY$\Sigma$:
As is shown in JPS87, the following relations hold:

\begin{itemize}
\item One has LY$\Sigma \subset$ SS$\Sigma$.
\item For each $i$ with $1\leq i\leq\ell$, there is an ergodic measure
$\mu$ with support in the spectral interval $[a_i,b_i]$.

\item Assume that the invariant set $\mathcal{M}$ is ``dynamically
connected'', that is, $\mathcal{M}$ cannot be written as the union
of two disjoint, nonempty, closed, invariant sets. Then the
following holds:
For $b_1$, the largest value in SS$\Sigma$, there is an
ergodic measure $\mu$  with the property that
$\lambda=b_1$ is a strong Lyapunov exponent.
It follows that $\lambda_1=b_1$ is a Lyapunov exponent
for some ergodic measure $\mu$ on $\mathcal{M}$,
and consequently there is a unit vector $x_1$ in $W_1(\theta)$ 
with the property that $x_1$ is a red vector.

\item The previous item is valid for any of the endpoints
$\{a_1,\cdots,a_k; b_1,\cdots b_k\}$, but the related
ergodic measures may differ.
\end{itemize}

It can happen, as in the case with the CY92 model, that there is a unique red vector. 
Furthermore, in this case, as is noted in Section \ref{examples}, one
has $\text{dim }V_1(\theta)=1$.
Hence one has $W_1(\theta)=V_1(\theta)$, and there is a unique
red vector in the projective flow.
Moreover, due to the exponential dichotomies occurring in the
CY92 model, the red vector is robust, and it varies continuously
with small changes in the model.

On the other hand, when $\text{dim }W_1(\theta)=1$, there is 
always a unique
red vector. If in addition, one has
$\text{dim }V_1(\theta) \geq 2$, then the red vector may be only
measurable and not continuous. In short, the red vector need not be robust.

\section{Continuum Limits of the BV and EBV Algorithms}
\label{interpret}
We now elucidate further the relationships between the BV and EBV algorithms and  
the dynamics of the underlying system.  In particular, we now formalize the connections between these 
algorithms and the solutions of the linear tangent equation, i.e., the finite-time Lyapunov vectors.
We are interested in the behavior as the step size $\delta t$ goes to $0$. 
To accomplish these aims we revert to a continuum formulation and for simplicity, assume that the vector
$y(t)$ in \eqref{EQ1} is a member of the Euclidean space $\mathbb{R}^K$. 
Since energy norms are used in many geophysical fluid mechanics problems, 
we will take the usual $l^2$-norm. We denote the norm and inner product by $\|\cdot\|$
and $(\cdot,\cdot)$, respectively, with $\|\cdot\|^2 = (\cdot,\cdot)$

The notation in this chapter differs from what has been set
in the rest of the paper in minor ways, like the letters representing the functions. This is so to 
emphasize that unlike what one computes in practice, the dynamical systems here are continuous in time.
However, everything is clearly explained to avoid ambiguities without cluttering the presentations with
 technical details. 

We stress that the term \emph{continuum limit} refers to the rescaling time (sometimes called a cycle),
 not to the numerical integration time step. In a numerical context, this corresponds to a strategy where
both integration time steps and the rescaling times are small.
While this has no importance for linear systems, it leads to different outcomes when
applied to nonlinear systems even for quite small perturbations amplitudes. 

We first recall the system \eqref{EQ1}:
\begin{align} \label{EQ1bb}
\begin{aligned}
\frac{dy}{dt} &= G(y), \quad t > t_0,\\
y(0) &= y_0.
\end{aligned}
\end{align}
Our first goal below is to obtain the formula
\begin{align}\label{eq:ctsbreeding}
W(T;\varepsilon)  &=
W(t_0;\varepsilon)+\int_{t_0}^{T}\left[G_{y}^{\varepsilon}\left(W(t;\varepsilon),t\right)
                                       -\left(G_{y}^{\varepsilon}\left(W(t;\varepsilon),t\right),W(t;\varepsilon)\right)W(t;\varepsilon)\right]dt,
\end{align}
where  for all $t\geq t_0$,
\[
W(t_0;\varepsilon)=W_0; \quad \|W_0\|=1;\quad G_{y}^{\varepsilon}\left(W(t;\varepsilon),t\right)=\frac{1}{\varepsilon}G_{y}\left(\varepsilon W(t;\varepsilon),t\right),
\]
and $ \displaystyle G_{y}(\delta y,t)=G(y+\delta y)-G(y)$ for the limiting case $\delta t\to 0$, $t_0\le t \le T$. 
The vector $W(t;\varepsilon)$ corresponds to a continuously rescaled bred vector $V(t)$ at amplitude $\varepsilon$ and 
with initial perturbation $V(t_0) = \varepsilon W_0$. ($W$ has the same direction as $V$, but it has amplitude $1$). 
We will streamline the presentation by skipping some of the techical details in derivation,
and we will assume at the outset that $G$ in \eqref{EQ1bb} has the necessary differentiability properties to make all 
the mathematical steps rigorous.

Let $y$ and $y+\delta y$ be two solutions of \eqref{EQ1bb} that satisfy the 
initial conditions $y(t_0)=y_0$ and $(y+\delta y)(t_0) = y_0+\delta y_0$.
Then $\delta y(t)$ is a solution of 
\begin{equation}
\frac{d}{dt}\delta y=\left[G(y+\delta y)-G(y)\right],\ \delta y(t_0)=\delta y_{0}.
\label{eq:deltafevol}
\end{equation}
Even if $G$ is autonomous, the resulting equation \eqref{eq:deltafevol} for $\delta y$ is non-autonomous.
Integrating \eqref{eq:deltafevol} for $t\ge t_0$, we obtain
\begin{equation} \label{eq:deltayintegr}
\begin{aligned}
\delta y(t_{0}+\delta t) 
  &=\left([y(t_{0})+\delta y(t_{0})]+\int_{t_{0}}^{t_{0}+\delta t}G(y+\delta y)\, dt\right)\\
 &\qquad \qquad \qquad\qquad \qquad - \left(y(t_{0})+\int_{t_{0}}^{t_{0}+\delta t}G(y)\, dt\right).
\end{aligned}
\end{equation}
We next define $\varepsilon = \|\delta y(t_0)\|$, and a family of vectors $v(t)$ on a sphere centered at zero with 
radius $\varepsilon $ in $\mathbb{R}^K$ by \ 
\[
  v(t)=\frac{\delta y(t)}{\|\delta y(t)\|}\|\delta y(0)\|.
\]

Assuming that $\delta y$ is bounded away from $0$, $v(t)$ obeys the evolution equation: 
\begin{align}
\frac{1}{\|\delta y(0)\|}\frac{dv}{dt} 
 & =\frac{1}{\|\delta y(t)\|^{2}}\left[\left(\frac{d}{dt}\delta y\right)\|\delta y(t)\|-\delta y(t)\frac{d}{dt}\|\delta y(t)\|\right]\nonumber \\
 & =\frac{1}{\|\delta y(t)\|^{2}}\left[G_{y}(\delta y,t)\|\delta y(t)\|-\frac{1}{2}\frac{\delta y(t)}{\|\delta y(t)\|}\cdot\frac{d}{dt}\|\delta y\|^{2}\right],\label{eq:uhat1}
\end{align}
which makes explicit the fundamental dependence of the time evolution of $v$ on the norm.
Taking the scalar product of (\ref{eq:deltafevol}) with $\delta y(t)$, we also have
\[
\frac{1}{2}\frac{d}{dt}\|\delta y\|^{2}=(G_{y}(\delta y,t),\delta y(t)).
\]
Substituting this relation in (\ref{eq:uhat1}) gives after some simplifications,
\begin{align}
\frac{1}{\|\delta y(0)\|}\frac{dv}{dt} 
 & =\frac{G_{y}(\delta y,t)}{\|\delta y(t)\|}-\left(\frac{G_{y}(\delta y,t)}{\|\delta y(t)\|},\frac{v(t)}{\|\delta y(0)\|}\right)\frac{v(t)}{\|\delta y(0)\|}.
\label{bredctsdiff}
\end{align}
We integrate (\ref{bredctsdiff}) independently  on  successive intervals 
$$[t_0,t_0+\delta t),\,[t_0+\delta t, t_0+ 2\delta t),
\ldots,\, [t_0+(N-1)\delta t, T)$$, where $T = t_0+ N \delta t$ 
and  denote the solution of \eqref{bredctsdiff} on the interval $I_k = [t_0+(k-1) \,\delta t, 
t_0+k\,\delta t)$ by $v_k(t)$. 
This is a sytem of $N$ integral equations  with $N$ free parameters (the integration constants) $v_k^0, k=1,\ldots, N$:
\begin{align}
v_k(t) & = v_k^0 +\|v_k^0\|\int_{(k - 1)\delta t}^{t}\left(\frac{G_{y}(\delta y_k,\tau))}{\|\delta y_k(\tau)\|}-\left(\frac{G_{y}(\delta y_k,\tau)}{\|\delta y_k(\tau)\|},\frac{v_k(\tau)}{\|v_k^0\|}\right)\frac{v_k(\tau)}{\|v_k^0\|}\right)d\tau, \nonumber \\
 &\qquad \qquad t_0+ (k - 1)\delta t \le t <t_0+k\,\delta t,  \quad k = 1,\ldots, N,
\label{eq:main}
\end{align}
where $\delta y_k$ satisfies the \eqref{eq:deltayintegr} with the initial condition
$v_k^0$ at $t=t_0+ (k - 1)\delta t $, i.e., the left-hand boundary of the interval $I_k$. 
We observe that, at $t = t_0+ (k - 1)\delta t$, $v_k(t)$ 
is an approximation of the discrete bred vector with short rescaling time step $\delta t$.
Hence, we take it as the basic approximation for the continuously rescaled bred vector we seek.
Note that the integrand in \eqref{eq:main} is simply the component of the vector 
$\frac{G_{y}(\delta y,t)}{\|\delta y_k(t)\|}$ perpendicular to $v_k(t)$ in $\mathbb{R}^K$ at time $t$. 

We take 
\begin{equation}
v_1^0 = \delta y (t_0), \qquad 
v_k^0 = 
\lim_{t\to t_0+(k - 1)\delta t^-} v_{k-1}(t), \quad 1\le k\le N,
\label{bvctsic}
\end{equation}
assuming the limits exists, and  extend the domain of definition of $v_k$ by setting
\ $v_k(t) = 0 \text{ if } t\not\in I_k$.
Similarly, we extend the domain of definition of $\delta y_k$. We then let \
$V_N = v_1 +\cdots +v_N,$
and  note that $V_N$ is continuous in time on $[t_0, t_0 + N\delta t]$ (extended to $t=t_0 + N\delta t$ by continuity) and piecewise differentiable. 
We define $\delta Y_N$ analogously as the sum of
extended $\delta y_k$'s.
Then $$V = \lim_{N\to\infty}V_N = \lim_{\delta t\to 0}V_N$$ exists and 
is continuous in time, for instance,
when $G$ has the necessary differentiability
properties so that the sequence $V_N$, possibly after passing to a subsequence, converges uniformly on compact time intervals, 
as $N\to \infty$ (or, equivalently, as $\delta t\to 0$). 
Since $V_N$ coincides with the right-hand side limit of $\delta Y_N$ at every grid point, 
the difference between $V_N$ and $\delta Y_N$ goes to zero as $N\to \infty$ in an appropriate norm that is allowed by how 
smooth $G$ is. When $G$ is continuously differentiable, for example, this convergence 
would be uniform on compact time intervals.
Thus, by passing to the limit $N\to \infty$ in \eqref{eq:main}, we obtain the following representation formula
for $V$: 
\begin{equation}
V=V(t_0)+\int_{t_0}^{T}\left(G_{y}(V(t),t)-\left(G_{y}(V(t),t),\frac{V(t)}{\|V(t)\|}\right)\frac{V(t)}{\|V(t)\|}\right)dt.
\label{veq}
\end{equation}
Under the regularity assumptions above on $G$, the integrand is
continuous and hence $V$ is a mild solution of an associated
differential equations. Therefore, we can bootstrap and prove the further regularity of $V$. Note that, by construction,
\[
\varepsilon=\|V(t_0)\|=\|V(t)\|.
\]
Therefore, Equation \eqref{veq} is equivalent to \eqref{eq:ctsbreeding}, if we take $\delta y_{0} = \varepsilon W_0$.

An immediate consequence of \eqref{eq:ctsbreeding} is the following. When $G$ is sufficiently regular,
$W(t;\varepsilon)$ converges uniformly on $[t_0, T]$ as $\varepsilon\to 0$. Let us denote the pointwise limit
by $W(t)$. Then, by \eqref{eq:ctsbreeding}, and the observation
\[
   \lim_{\varepsilon\to0}G_{y}^{\varepsilon}\left(W(t),t\right)
   =\lim_{\varepsilon\to0}\frac{G(y+\varepsilon W(t))-G(y)}{\varepsilon}
   =D_{W(t)}G(y),
\]
where the right-hand-side is the directional derivative of $G$ in the $W(t)$-direction, 
it follows that $W(t)$ satisfies the differential equation
\begin{subequations}
\label{eq:crtlm}
\begin{equation} \label{eq:crtlm.a}
\frac{dW}{dt} = A(t)W - (A(t)W, W)W, 
\end{equation}
where 
$$
A(t) = D_yG \vert_{y=y(t)},
$$
and 
\begin{equation}
W(t_0) = W_0, \quad 
 W\in \mathbb{R}^K. \label{eq:crtlm.b}
\end{equation}
\end{subequations}
We recall the Linear Tangent Equation for the problem \eqref{EQ1bb}: 
\begin{subequations}
\label{eq:lineartangenteqn}
\begin{equation} \label{eq:lineartangenteqn.a}
\frac{dX}{dt} = A(t)X,\quad A(t) = D_yG\vert_{y=y(t)},
\end{equation}
\begin{equation} \label{eq:lineartangenteqn.b}
X(t_0) = W_0, \quad  X\in \mathbb{R}^K,
\end{equation}
\end{subequations}
where $D_yG$ is the Jacobian matrix of the map $G(y)$, and $y(t)$ is the solution of \eqref{EQ1bb}.\ 
As in Section \ref{problem}, $U(t,s)$ denotes the solution operator of
\eqref{eq:lineartangenteqn}, taking the solution at time $s$ to the
solution and time $t$. With a slight abuse of notation, we write $U(t) :=
U(t,t_0)$, so that in particular the solution $X(t)$ of
\eqref{eq:lineartangenteqn} at time $t$ is simply $U(t)W_0$. We recall also that $U$ is a semiflow, {\it i.e.},  
\begin{equation} \label{eq:semiflow}
 U(t+s)= U(t+s,t)\, U(t), \qquad \forall \, \, t, \, s \in \R. 
\end{equation} 

What is the relation between \eqref{eq:crtlm} and \eqref{eq:lineartangenteqn}?
It can be easily seen that rescaling $X(t)$ to unit length 
and restarting the integration of \eqref{eq:lineartangenteqn} with $X(t)/\|X(t)\|$ at any $t_m\in[t_0,T]$
does not alter value of $X(t)/\|X(t)\|$ for $t\ge t_m$. By induction, this follows for any finite number of
rescaling-restarting cycles. This is because of the linearity of \eqref{eq:lineartangenteqn}.
We leave the details to the reader. By a direct approximation argument, 
or by a similar argument that led to \eqref{eq:crtlm}, we obtain
\[
	\frac{X(t)}{\|X(t)\|} = W(t), \quad t\in[t_0, T].
\]
Thus, we have just established that, the continuously rescaled bred vector $W(t;\varepsilon)$, i.e., the solution of 
\eqref{eq:ctsbreeding}, converges
to the corresponding solution (rescaled to size $1$) of the tangent linear equation \eqref{eq:lineartangenteqn} 
with the same initial data, i.e., to the solution of \eqref{eq:crtlm},
as $\varepsilon\to 0$ uniformly on the compact set $[t_0,T]$, assuming that G is smooth enough. 
Since the solutions of \eqref{eq:crtlm} have constant magnitude $1$, it is the linear analog of \eqref{eq:ctsbreeding} in
terms of continuous rescaling. 

While continuous rescaling is inconsequential for linear equations, this is not the case for the nonlinear ones.
However, when the initial amplitude is very small, the results above indicate that one might be able to rescale less often 
and still get comparable results. This is simply because it takes longer for nonlinear effects to start to dominate the picture. 

\subsection{Continuum Limit for EBV}

We discuss the continuum limit for the EBV algorithm briefly, by confining ourselves to a short, heuristic description.
Writing a system of differential equation for the EBV algorithm is not as straightforward as  for the BV algorithm.
The perturbations $\delta y(\iota)$ in the ensemble that grow fastest at the largest amplitude (before rescaling) 
at time $t$ will  obey the equation \eqref{eq:ctsbreeding} as long as they are the top contenders. 
All the other perturbations will satisfy the following differential equation:
\begin{align}\label{eq:ctsebv}
\frac{d(\delta y(\iota))}{dt} &= G_y (\delta y(\iota),t) - \frac{\delta y(\iota)}{\varepsilon }\frac{dh(t)}{dt},
\end{align}
where $\iota \in\BI$, and \
$
dh(t)/dt
$
is the maximum growth rate of the norm among  all ensemble members with the maximum amplitude $\varepsilon$ at time $t$.
Assuming enough smoothness on $h(t)$, \eqref{eq:ctsebv} can be heuristically obtained
 by taking the right time derivative  of the vector 
\ $\displaystyle
\delta y(\iota)(t)\frac{\varepsilon}{h(t)},
$
where $h(\tau), \tau\ge t$ is the maximum norm of the ensemble members at time $\tau$ 
in absence of any rescaling after time $t$ (If Equation \ref{eq:deltafevol} were in charge). 
As $\varepsilon$ approaches zero for  time $t$ fixed, all  ensemble members point in directions 
along which \eqref{eq:ctsebv} approaches the linear tangent equation, since 
$dh(t)/dt \to 0$  by continuity of $G_y$ at zero with $G_y(0,t) = 0$. 
A detailed treatment complete with the technical aspects will be presented elsewhere.

\subsection{Tuning the Maximum Amplitude}

We note that the convergence to the linear tangent equation is expected to be quicker for 
the EBV compared to the  BV for most of
the ensemble members. This is manifested in the equations, and later verified numerically in the next chapter.
The acceleration is due to the inherent size ordering in the EBV algorithm: 
information on the scales where linear tangent equation
dominates tends to be preserved in a robust manner against the changes in the parameter $\varepsilon$.
On the other hand, at finite amplitude away from $0$, the rescaling rule in the BV algorithm might create
or prolong the life of certain instabilities due to the nonlinear effects. The algorithm output needs to be examined
independently to check whether these vectors are in fact relevant to the dynamics or just artefacts of the rescaling strategy.
We believe that the EBV algorithm is more resistant and robust in this regard. 
The parameter $\varepsilon$ is tuned under different considerations in the EBV and BV algorithms.

\subsection{Separation of Scales}

In nonlinear systems,
it is possible that a perturbation will not grow very rapidly  
in the zone of perturbations with size near $\varepsilon$, 
but instead, a multiple of this perturbation
can be dominant among the smaller perturbations and might grow quite fast there. 
It is also possible that the dominant vectors in different amplitude zones will
be close in structure, but very different in growth characteristics. These theoretical considerations
are realized in the Lorenz63 system quite strikingly. See Figure \ref{difflorenz2b} and its discussion, 
especially the recurring patterns of ``spears'' therein.
This can be seen very easily from the following relation:
$$\frac{\|V(t_n)\|}{\|V(t_{n-1})\|} = \bR^{\text{min}}_n\frac{\|\bar{V}(t_n)\|}{\|V(t_{n-1})\|},$$
where ${\bf{R}}_n^{\text{min}}$ is defined similarly to \eqref{r-bold}, $V(t)$ denotes a EBV$(t)$ member, and
$\bar{V}(t)$ denotes the same vector right before the rescaling.
In a nonlinear system, 
this ratio can be larger than $1$,
for some range of perturbation sizes smaller than $\varepsilon$. This is the main mechanism
that creates the vector zones of different magnitude with zonal growth characteristics, or a separation of scales.

Another case in point will be  presented in Section 
\ref{examples} for the CY92 model. Even at  late times 
and  small amplitudes, we can see perturbations surviving the rescaling strategy and 
they resemble what one would get
from the usual BV algorithm at those sizes (and the finite-time Lyapunov vectors) 
very closely, whereas the dominating perturbation of size $\varepsilon$
resembles a particular BV of size $\varepsilon$. This example actually illustrates more. 
There are members of the ensemble in small magnitudes that
are slow in aligning with the dominant directions. 

\section{Applications of the Bred Vector Algorithms}
\label{examples}

We will be comparing the BV and the EBV on two problems: The Lorenz equations, 
or Lorenz63 (see \cite{lorenz63}), and a dissipative and forced nonlinear 
partial differential equation that arises in modeling  the 
 thermohaline circulation (see \cite{cy92}). The latter will be denoted as the CY92. It 
 is a Cahn-Hilliard  equation and it will shown to have an inertial manifold.
We will also have occasion to compare the BV and EBV results to the finite-time
Lyapunov vector outcomes.

Throughout we will use an explicit fourth order Runge-Kutta time marching scheme, for
the calculation of the base solutions as well as for the calculations of the BV, EBV, and finite-time
Lyapunov vectors.

\subsection{The Lorenz63 Model}

The  finite-dimensional, nonlinear Lorenz63 model has often been used as benchmark  
for testing sensitivity and, in particular, 
as a test problem for BV (see  for example \cite{evans04}).

Let $X=(X_1,X_2,X_3)\in \BR^3$ and let 
\begin{equation}
\label{L2}
A=\left(\begin{array}{ccc}
	-\sigma &\sigma & 0 \\
	r & -1 & 0 \\
	0 & 0 & -b
\end{array}\right),
\, N(X)=
\left(\begin{array}{c} 0 \\ -X_1X_3 \\ X_1 X_2
	\end{array}\right),
\, DN(X)=\frac{\partial}{\partial X} N(X).
\end{equation}
$DN(X)$ is a $3\times 3$ matrix-valued function.
The Lorenz model is described by the solutions of
the nonlinear system:
\begin{equation}
\label{L1}
\partial_t X\,=\,A\,X+N(X).
\end{equation}
The associated Tangent Linear Model  
is the skew product system
\begin{equation}
\label{L3}
\begin{aligned}
\partial_t X\,&=\,A\,X+N(X) \\
\partial_t U\,&=\,\left(A+DN(X)\right)\,U,
\end{aligned}
\end{equation}
where $U=(U_1,U_2,U_3)\in\BR^3$.
The $U$-equation in \eqref{L3}, which is a linear equation,
is of special interest to us.

We will use the notation for the Tangent Linear Model introduced  in Section \ref{sec.LV}. With the initial condition
$\theta=X_0\in\BR^3$, we let
$\theta\cdot t=S(\theta,t)=X(t)$ denote the solution of the
nonlinear equations \eqref{L1} that satisfies $S(\theta,0)=\theta$.
In this study we set $r = 28, b = 8/3,  \sigma = 10$. It is well-known that for these
values of the parameters the Lorenz63 model has a chaotic global attractor.
For the non-autonomous $U$-equation; 
$\partial_t U=\left(A+DN(\theta\cdot t)\right)\,U$,
we let
\begin{equation}
\label{U(t)}
U(t)\,=\,U(\theta,t)\,U_0,\quad\text{ denote the solution operator,}
\end{equation}
where $U_0\in\BR^3$, and $U(0)=U_0$.

We will use
the term ``attractor'' to refer to a compact, invariant set $\fma$
that attracts a neighborhood of itself. As a result, $\fma$ is
Lyapunov stable, as a set. Consequently, for $0<\varepsilon<1$,
there is a family of $\varepsilon$-neighborhoods, 
$N_{\varepsilon}$, of $\fma$,
where each neighborhood is positively
invariant and $\fma=\cap_{\varepsilon>0}N_{\varepsilon}$.
When we write that $\theta$ is near $\fma$, we mean that
$\theta\in N_{\varepsilon}$, for some small $\varepsilon>0$.
See Chapter 2 in \cite{SY02}, for a history 
of this concept and more information.

By using a somewhat different --but equivalent formulation-- Toth \&
Kalnay  have suggested that the time evolution
$BV(t)$ is a good approximation of the tangent linear
solution $U(t)$, over bounded time-intervals.
As a consequence of the continuum limit theory in Section \ref{interpret},
we see that this perceptive observation has a solid mathematical
basis, provided that the time-step $\delta t$ is small and the perturbation amplitude is small, as well.

In order to find a numerical validation of the continuum limit
theory described in Section \ref{interpret}, we will calculate the distance
\begin{equation}
\label{D(t)}
D(t,\delta t)\,:= \,d_{\mathrm{pr}}\,(\left[BV(t)\right],\,\left[U(t)\right]),
\end{equation}
where $d_{\mathrm{pr}}$ is the projective metric on $\fma\times\mathcal{P}^{K-1}$
(see \eqref{d-pr}), and $BV(t)$ is a BV at time $t$. 
For this calculation, we assume that both $BV(t)$ and $U(t)$
satisfy the same initial condition $(y_0, \delta\mathcal{Y}_0)$
at $t=0$, and that $y_0=\theta$ is near the attractor $\fma$.
We also use identical ensembles of perturbation vectors for
both the BV and the EBV algorithms.
We then fix $T>0$, and we examine the distances $D(T,\delta t)$
(in the projective space), for different choices of $\delta t$.
Our goal is to show that $D(T,\delta t)$ becomes smaller,
as $\delta t$ gets smaller.
The max and min values, for both the $BV$ and the $EBV$  
algorithms, are reported in Table \ref{tab1}.
For 
the calculations used to generate the data in Table \ref{tab1},
we had set the initial conditions for both $BV(t)$ and $U(t)$,
at $t=0$, so that
$X=0.5688$, $Y=0.4694$, $Z=0.0119$, 
and $\delta\mathcal{Y}_0=[1,1,0]$.
For the ensemble, we took the 3-fold Cartesian product 
of the set
$$
\{-1,-0.75,-0.5,-0.25,0,0.25,0.5,0.75,1\}
$$
in $\BR^1$, and projected this product
onto the sphere of radius $0.1$, centered at the origin, in $\BR^3$. 
By eliminating the repetitions introduced with this projection,
one obtains an ensemble of 584 distinct points in $\BR^3$.
(Note that the initial conditions are ''near'' the Lorenz attractor.)
The trajectory for the nonlinear problem
was computed with a step size $dt = 0.0001$, so that any observed
variations are only due to the differences
in the algorithms and the relative effects on the perturbations.
In  Table \ref{tab1}, we fixed $T=2$ and made two choices for $\delta t$, 
namely $\delta t=0.004$ and $0.001$. The results compare the outcomes
of the  $BV$ and the $EBV$ algorithms. As noted in Section 3, both algorithms $EBV(t)$ and $BV(t)$
approximate solutions of the tangent linear equation,
as $\delta t$ goes to 0. By using the projective metric 
one can compare the rates of convergence in terms of this metric.

The maximal values of $D(T,\delta t)$,
for both the BV and EBV algorithms, are essentially the same 
for both choices of $\delta t$, and the minimal values are
essentially the same for $\delta t = 0.004$. 
When one moves from $\delta t=0.004$ to $\delta t=0.001$,
both minimal values decrease, as expected. However the drop
in the minimal value for the EBV algorithm is substantially
larger than the drop for the BV algorithm.
By using the perturbation corresponding to the minimal
drop in $D(T,\delta t)$, for each of the algorithms, one arrives
at the {\it best approximation \,} given by for the given
algorithm. Clearly the min BV and min EBV columns in Table \ref{tab1}
shows that the EBV algorithm yields a better approximation
than the BV algorithm.

\begin{table} 
\caption{{\it Maximum and  minimum values of $D(T,\delta t)$ with $T=2$, for two $\delta t$. See (\ref{D(t)}).}}
\begin{tabular}[t]{|r||c|c||c|c|} \hline
\quad $\delta t$ \quad & max BV & min BV &
	max EBV & min EBV \\
\hline\hline 
0.004 & $1.40 \times 10^{-1}$ & $5.26 \times 10^{-4}$ &
	$1.01 \times 10^{-1}$ & $1.10 \times 10^{-3}$ \\
0.001 & $7.77 \times 10^{-2}$ & $1.03 \times 10^{-4}$  & 
	$3.48 \times 10^{-2}$ & $4.16 \times 10^{-5}$ \\
\hline
\end{tabular}
\label{tab1}
\end{table}

One can, of course, use different initial conditions
for the two solutions $BV(t)$ and $U(t)$ and/or larger
perturbations. However, one cannot expect to replicate the
results seen in Table \ref{tab1} in that case. 
First, there is a transient phase, which ends when the
two solutions are ''near''   the attractor.
Even if this transient phase is short, one still has a problem.
While the attractor is stable, as a set, one still has a problem
because the flow on the attractor is generally not
Lyapunov stable.
Essentially all pairs of nearby
orbits may diverge over long time intervals.

\begin{figure}
\centering
\includegraphics[scale=1]{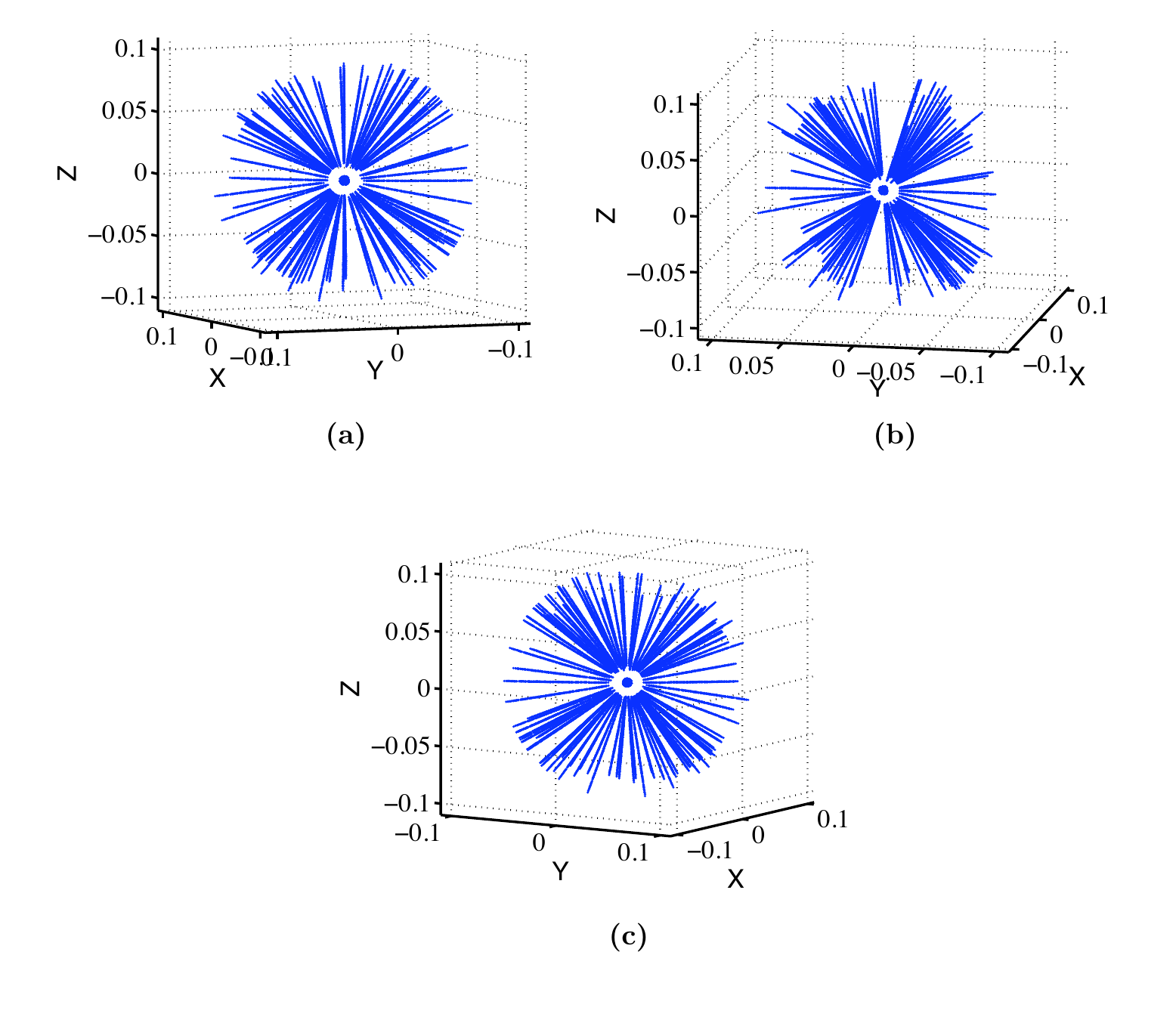}
\caption{{\it Viewed from three different angles, (a)-(c). EBVs, for times $24$ through $30$, taken at $0.1$ time intervals, for Lorenz63. Parameters and conditions are those used to generate Table \ref{tab1}.}}
\label{zoom1}
\end{figure}

\begin{figure}
\centering
\includegraphics[scale=1]{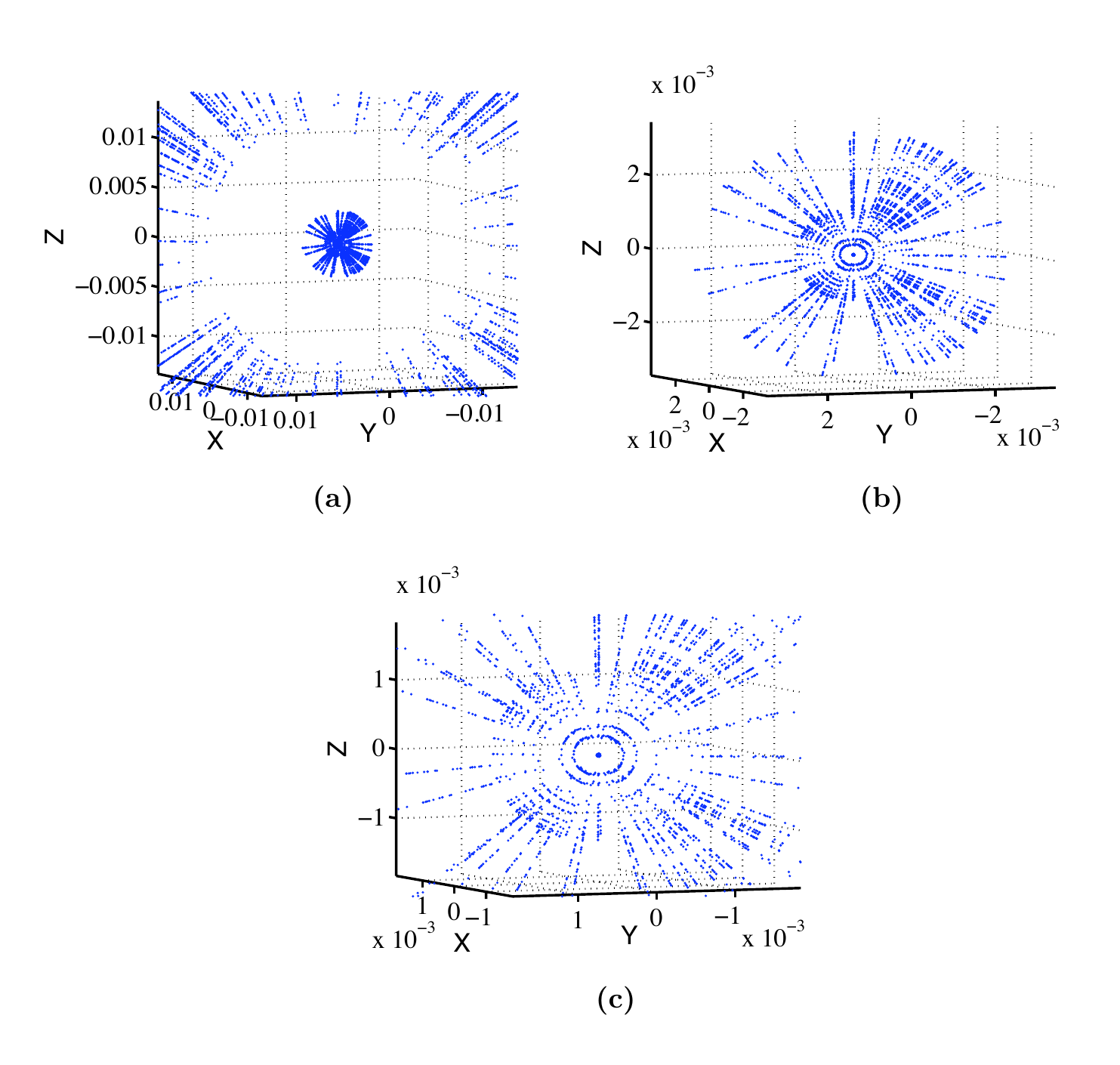}
\caption{{\it Close-up views  of Figure \ref{zoom1}a: (a) zoomed in 8 times; 
(b) zoomed in 32 times; (c) zoomed in 60 times. Note the preservation of structure even
as we zoom in.}}
\label{zooma}
\end{figure}

An interesting outcome of the use of the EBV algorithm on
Lorenz63 is that it exposes the 
fractal behavior of the Lorenz attractor.
We computed a 584-member EBV, using the same parameter values, 
initial conditions and base trajectory as was used in generating Table
\ref{tab1}, for the $\delta t = 0.001$ case. We constructed plots by merging
all the vectors between time $24$ and $30$, at intervals of $dt=0.1$.
In  that time, at any given instant, one only observes a couple of
members reaching  highest amplitude.  By rotating the same plot about the $X_3=Z$ axis, one obtains three
views of the EBV's  shown in Figure \ref{zoom1}.
Zooming into Figure \ref{zoom1}a by a factor of 8, 32, and 60, respectively, we obtain Figure \ref{zooma}.
The fact that one observes similar patterns at several levels of magnification
meets the requirement of `fractal behavior' that was introduced by
\cite{BM77}.

\begin{figure}
\centering 
\includegraphics[height=3.in]{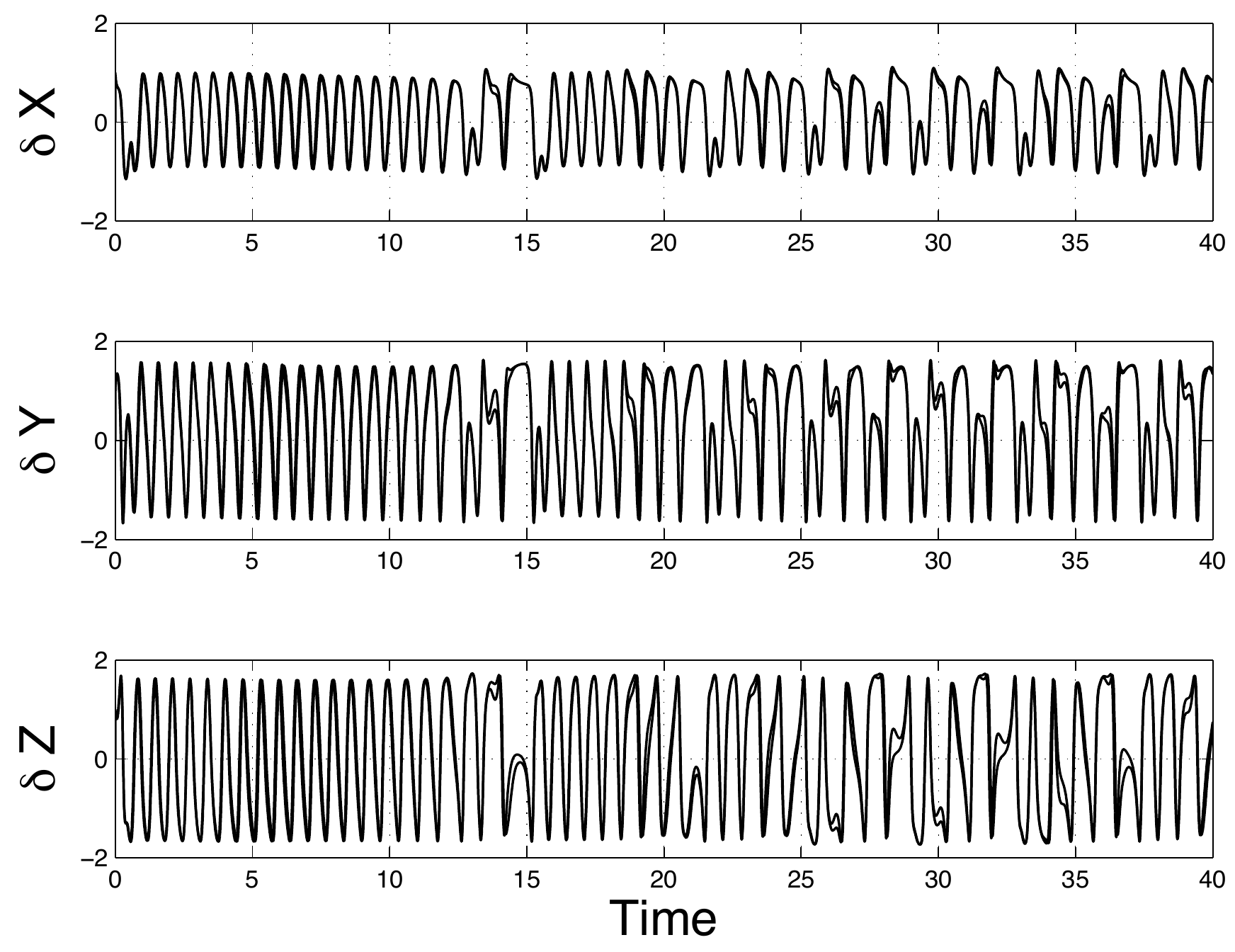}
 \caption{{\it  The three components of  BV and of the  finite-time Lyapunov vectors, as a function of time, for the Lorenz63 system. 
 (For parameters and initial conditions, see text).  The vectors are nearly coincidental over the whole time span. The perturbation size is $1$, in all of the components. 
For our choice of initial conditions the vectors exhibit more temporal regularity at earlier times. This transient behavior disappears at $t=13$, approximately.}}
 \label{reglorenz2}
\end{figure}

Figure \ref{reglorenz2} shows the time evolution of a sample BV and
the  corresponding finite-time Lyapunov vector with the same initial perturbation. The results were obtained                                           
with a time step of $\delta t = 0.005$. The computations of the BVs and finite 
time Lyapunov vectors                                                           
were performed with a time interval equal to                                     
the dynamics time step. At $t=0$,                                               
$X=0.1493$, $Y=6.2575$, $Z=1.8407$.                                             
  The initial perturbation vector  $\delta {\cal Y}_0$  was $[1,1,1]$. As expected, small to moderate-sized perturbations 
  produces similar results in the finite-time Lyapunov calculation and the Bred Vector calculation.

\begin{figure}
\centering 
\includegraphics[height=3.in]{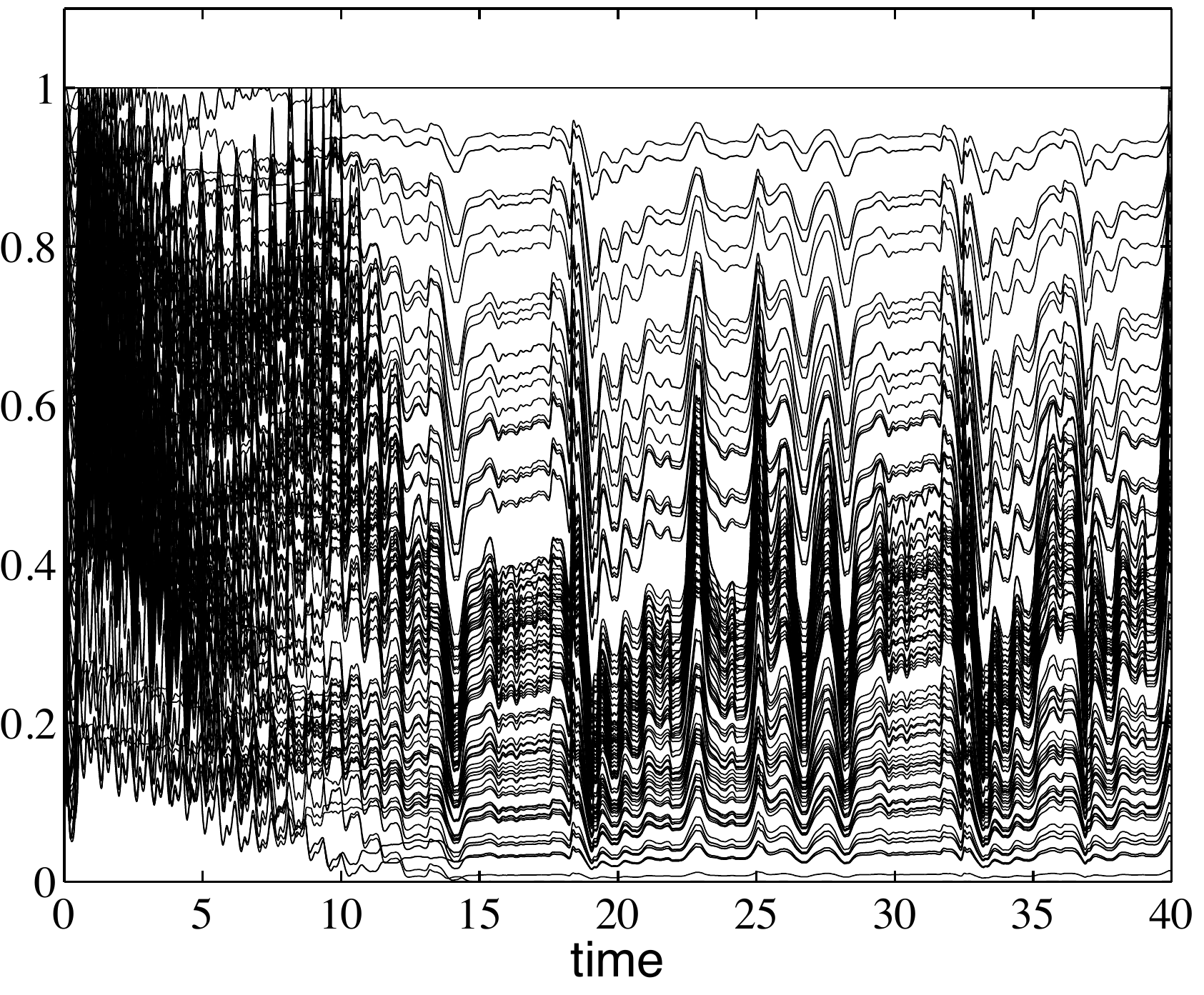}
 \caption{{\it 
 Plot depicting the
 evolution of the 2-norm of an ensemble of 98 EBV$(t)$. Same parameters and initial conditions, as in
 Figure \ref{reglorenz2}. 
 Initially, in the interval $[0,13]$ we note a very fast decay of some of the ensemble
 members, leading to a sorting in size, beyond that time. As described in the text, this outcome is one of the most
 distinguishing features of the EBVs, when compared to an ensemble of individual BV outcomes.}}
 \label{difflorenz2b}
\end{figure}

Figure   \ref{difflorenz2b} is devoted to the results of a calculation of
$EBV(t)$ alone, for the same case.
The figure depicts a rescaled time evolution of the norm of 98 distinct ensemble members in the EBV calculation. 
The rescaling is done with respect to the usual Euclidean $l^2$-norm.
The initial perturbations 
are made to sample a perturbation sphere of amplitude 1 about the initial conditions.
The figure highlights the rapid decay of many of the vectors and the eventual size-ordering that is inherent in the EBV algorithm.
For $0<t\leq 13$, the results in Figure \ref{difflorenz2b} describe the transient behavior and are rather chaotic.
However, for $t\geq13$ a very interesting pattern evolves: we see that the largest member
of the ensemble  takes on the value 1, for all $t\geq0$, 
which is expected. (This corresponds to the Lyapunov vector
with the largest exponent.) From the totality of the information on the ensemble, it is possible to 
extract the next two Lyapunov vectors, where the
exponents satisfy $\lambda_3\leq\lambda_2\leq\lambda_1$. However, this would require a deeper analysis,
 and it is deferred to a future work. In fact, viewed as 
a whole, the graphs corresponding to those smaller than 1 have very useful
structural information:
Notice the recurrent `spear-like' pattern, which occurs after $t=17$.
What is happening is that  
a group of `small' vectors in the attractor grow rapidly, as they
go around the horn in the Lorenz attractor. We propose that the rationale for this behavior is that the
nonlinear equations of motion temporarily
overwhelm the uniform rescaling rule for these small vectors.
For $t\geq 17$, we are beyond the transient zone, and we would not
expect such behavior to occur if the linear term $AX$ strongly dominates
the nonlinear term $N(X)$ in the vicinity of the attractor.
This is an excellent illustration of the fact that EBV algorithm preserves the
role of the nonlinear terms in the equations of motion.

This feature of the EBV can not be replicated by  BV,
even a BV with an ensemble of perturbations. The reason
for this is that the rescaling rule for the BV algorithm
forces all the perturbations to have the same norm,
for each $t\geq0$. Thus the Figure \ref{difflorenz2b} for the EBV would
be replaced by a figure for the BV, where all of the 
perturbations are plotted on the top line only.

\subsection{A Cahn-Hilliard Equation}
 \label{cy92}

\begin{figure}[ht]
\centering
\includegraphics[scale=1]{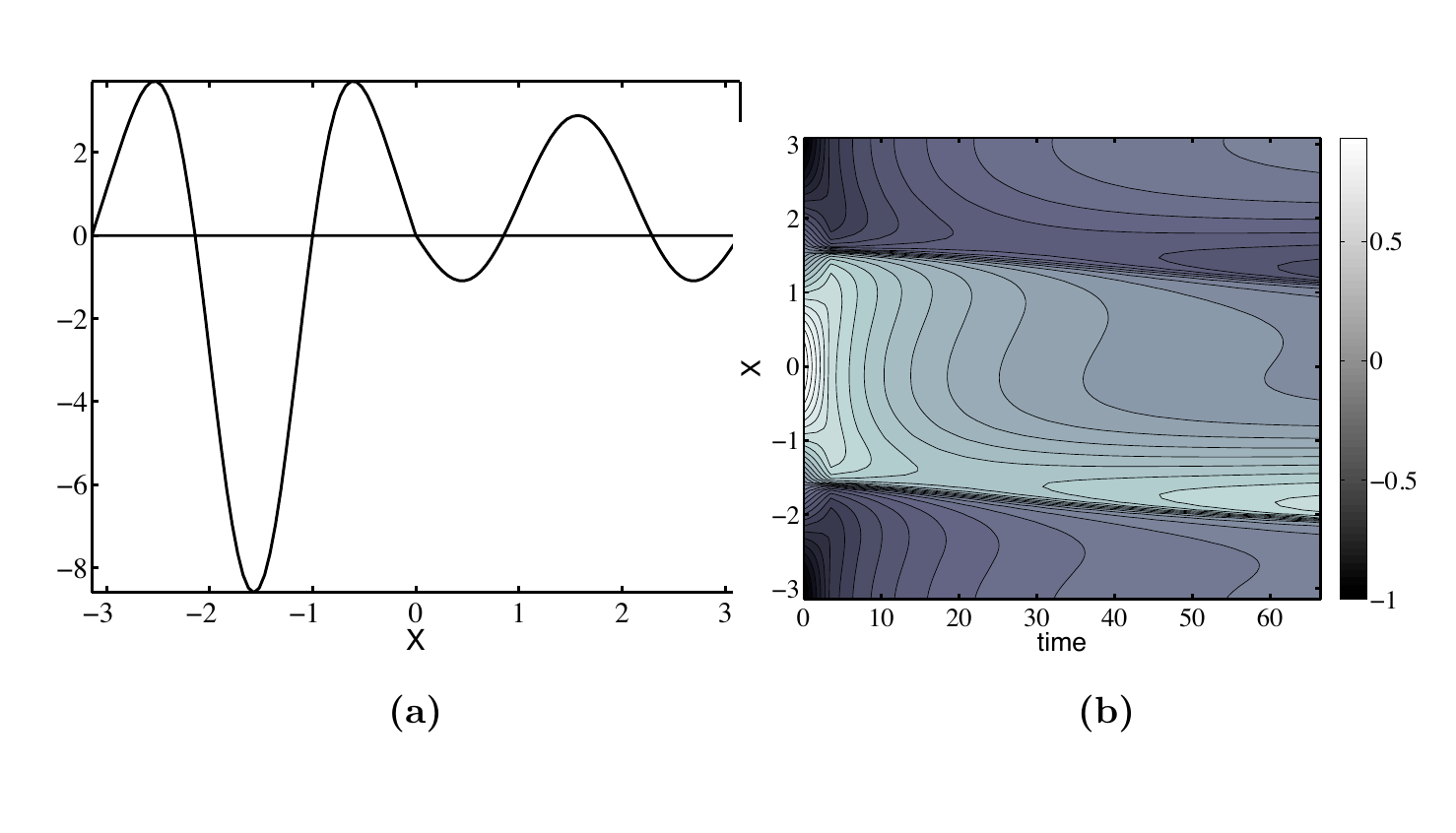}
\caption{{\it  (a) Forcing function $f_{xx}$, (\ref{ff}), used in the CY92 model;   the numerical solution to CY92 is shown
in (b). See Figure \ref{cycase2} for the BV algorithm and finite-time Lyapunov vector algorithm outcomes.}} \label{cycase}
\end{figure}

In their work on the thermohaline dynamics \cite{cy92} proposed a coupled model for the circulation, salinity, 
temperature, and density of the oceans, with atmospheric forcing. The crux of the model is
the partial differential equation for salinity:
The slow-time dynamics of the ocean salinity $S(x,t)$, zonally-averaged, and as a function of
latitude $x\in[-\pi,\pi]$ and time $t \ge 0$, is described by
\begin{eqnarray}
\frac{\partial S}{\partial t}&=& \alpha \frac{\partial^2}{\partial x^2} [f(x) + \mu S (S-\sin(x))^2 + S 
-\gamma \frac{\partial^2 S}{\partial x^2}], \quad  t>0, \nonumber \\
S(x,0) &=& S_0(x).
\label{CY60}
\end{eqnarray}
The equation is subject to zero-flux and zero-stress boundary conditions at the poles, however, we will be considering
periodic boundary conditions (for steady and periodic forcing $f(x)$ as well as equilibrium
solution $\sin(x)$ the conclusions that follow apply to the zero-flux
case). 
The positive parameter $\alpha$  affects the strength of the linear stability
of the model. We fix $\alpha=3.5 \times 10^{-3}$, 
$\gamma = 0.001$, and $\mu=\sqrt{10}$. The forcing $f(x)$ is a prescribed function that reflects balances
of evaporation and precipitation of freshwater; it can be symmetric, about the Equator ($x=0$), but is more typically,
asymmetric (see \cite{cy92eyink}). 
The second derivative of the forcing function, 
  will be chosen to be
\begin{eqnarray}
\partial_{xx} f(x) = \left\{ \begin{array}{c}
\frac{2}{27} (36-39 \mu^2 \cos^2 x-81 \cos^2 x
             + 8 \mu^2 + 25 \mu^2 \cos^4 x ) \sin x,  \\
             \mbox{for} \quad x \in [-\pi,0], \\
             \\
             -\frac{2}{9} (3 \mu^2 \cos^2 x - 3 - \mu^2) \sin x, \\
             \mbox{for} \quad x \in (0,\pi].
\end{array} \right.
\label{ff}
\end{eqnarray}
It is shown in Figure \ref{cycase}a. 
The initial condition chosen for this computation is $S_0(x) = \cos (x) $. The base solution obtained numerically is displayed in Figure \ref{cycase}b.

A great deal is known and can be said about the mathematical structure  of 
CY92 and its solutions. By letting $S=\partial_x u$ and $f=\partial_x F$ the CY92 can be
related to the 
 Cahn-Hilliard Equation (CHE) with forcing: 
 The general CHE, for $u$ above,  is
\begin{equation}
\label{CH1}
\partial_tu+\nu\triangle^2\,u\,=\,\triangle(g(u))+F
\qquad\text{for }x\in\Omega\text{ and }t\geq0,
\end{equation}
where $u=u(t,x)$ is a scalar field, 
$\nu>0$ is a constant, $\triangle$ is the Laplacian operator,
$\triangle^2$ is the bi-harmonic operator, and $g$ is a polynomial
of degree 3:
$g(u)=\sum_{j=1}^{3}a_j\,u^j$, with $a_3>0$.
The term $F=F(x)$ is the forcing function.
In general, the domain
 $\Omega$ may be an open bounded domain in 
the Euclidean space $\BR^m$,
with $1\leq m\leq 3$. However, we restrict our attention,
to the case where $m=1$ and $\Omega$ is the interval
$\Omega\,=\,(-\pi,\pi)$, with boundary $\Gamma=\{\pm\pi\}$.
As we will see, the multiplicity of the largest Lyapunov exponent
for this problem is 1.

For the analysis of the solutions of the CHE, one will use the
standard Sobolev spaces $H^j=H^j(\Omega)$, where $j\geq0$ is an 
integer and $H^0=L^2(\Omega)$.
As usual, the inner product and norm on $H^0$ is denoted by
$\langle \cdot,\cdot\rangle\,=\,\langle \cdot,\cdot\rangle_0$
and $\|\cdot\|\,=\,\|\cdot\|_0$.
For $j\geq0$ one uses
$$
\|u\|_{j+1}^2\\,=\,\|u\|_j^2+\sum_{|\alpha|=j+1}
	\int_\Omega |D^\alpha u|^2\,dx,
$$
where 
$$
D^\alpha=\frac{\partial^\alpha}{\partial x^\alpha},\quad
\alpha\geq0, 
$$
and
$$
\langle u,v\rangle_{j+1}\,=\,\langle u,v\rangle_j
	+\sum_{|\alpha|=j+1}\int_\Omega (D^\alpha u)\,(D^\alpha v)\,dx.
$$

The finite-time Lyapunov vector algorithm for the CY92 model is derived, by first rewriting the equation as
\begin{align*}
\partial_{t}S & =\alpha \partial_{xx}\left[\mu^{2}S\left(S-\eta\right)^{2}-rf(x)+S-\gamma^{2}\partial_{xx}S\right]=:F(S).
\end{align*}
We opt here to first linearize and then discretize. 
To obtain the  tangent linear equation, we need to find the linear map ${\cal L}$ such that
\[F(S+\delta S)-F(S)={\cal L}(\delta S)+o(\delta S)\text{ as }\|\delta S\|\to0,
\]
where $\| \cdot\|$ is a norm. Typically, this will be the norm in the
space where solutions live, such as the Sobolev space $H^1$ for
$S$, or dictated by physical considerations. In this
infinite-dimensional model, different norms are not necessarily equivalent.


First, we note that
\[\left\{ S+\delta S\right\} \left(\left\{ S+\delta S\right\} -\eta\right)^{2}-S\left(S-\eta\right)^{2}=\left(S-\eta\right)\left(3S-\eta\right)+o(\|\delta S\|).\]
We let $L$ be a finite-difference approximation to ${\cal L}$. 
The discrete tangent linear equation becomes the product of two matrices  $A B$,
where  $B$ is the matrix whose entries are found by discretizing  the operator
\[\mu^{2}\left(S-\eta\right)\left(3S-\eta\right)+Id-\gamma^{2}\partial_{xx},
\]
where $Id$ is the identity matrix, and $A$ is the matrix corresponding to the discretization of  $\alpha \partial_{xx}$. The finite-time Lyapunov algorithm  is obtained by
solving
\[\partial_t \delta \Theta = L \, \delta\Theta=AB \, \delta \Theta=A(B(\delta \Theta)),\]
for $t\in[0,T]$,
subject to some initial vector perturbation $\delta \Theta(0) = s$, where $\delta \Theta$ is the discretization of the perturbation $\delta S$. The perturbation is normalized to the norm of the initial 
perturbation at each $\delta t$, the time step of the explicit Runge-Kutta 4 time integration scheme
employed here. $T=70$.
 In the 
  simulations, $\delta t =0.01$.  We applied second-order centered finite
  differences in space, and  used 121 grid points, $-\pi=x_0, x_1,
  ...,x_{120}=\pi$.

 By construction the finite-time Lyapunov vectors are not amplitude sensitive, however the BVs are, thus  different amplitude perturbations will yield different BVs, in the case of a general nonlinear problem. This outcome has important practical implications, if one would like to use BV to either infer the structure in the field, or the degree of sensitivity of the outcomes to perturbations in initial conditions. A challenging problem could thus arise in the context of large-scale simulations: what is considered a large  structural change or a highly sensitive outcome is physics-dependent, perhaps even difficult to surmise quantitatively; the physics in question may not be fully understood and thus 
 a reasonable perturbation amplitude is simply guessed.

\begin{figure}[ht!]
\begin{center}
\includegraphics[scale=1]{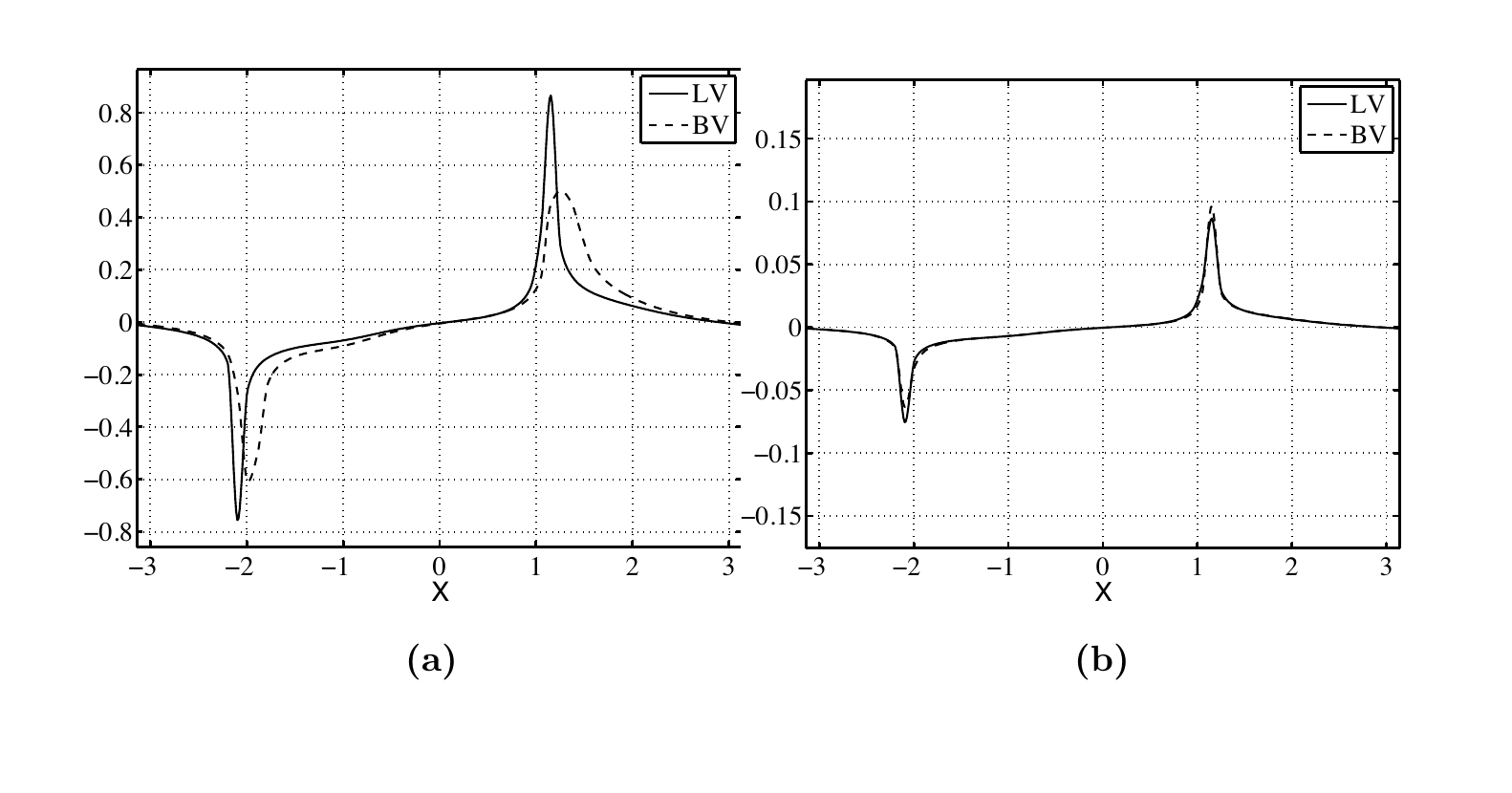}
\end{center} 
\caption{{\it Comparison of the finite time Lyapunov vectors and BVs. Effect of amplitude of perturbation on the 
 CY92 Model, with initial conditions $Y_0=\cos(x)$. The initial perturbation was $\epsilon \sin (x)$.
 Comparison of the finite-time Lyapunov vector and the BV outcomes at 
 $t=70$.  
  (a) Corresponds to $\epsilon=0.25$; (b) to $\epsilon=0.025$.}} 
\label{cycase2} 
\end{figure}

We ran the finite-time Lyapunov vector  and the BV algorithms, using the same initial condition,
forcing and perturbation. First, we examine the effect 
of the size of the amplitude of the perturbation on the outcomes: we
do so by keeping the shape of the perturbation 
fixed, changing only the overall amplitude.  Figure \ref{cycase2}a and b 
are plots of  the final BVs and finite-time Lyapunov vectors, corresponding to 2-norm 0.25 and 0.025
sized perturbations, respectively.  When the perturbations are 
small the BVs and the finite-time Lyapunov vectors are qualitatively consistent and, nearly so, quantitatively. 
The outcomes shown here are typical of the general case, that is, the qualitative and
quantitative disagreement grows with an increase in the 
amplitude of perturbations.

\begin{figure}
\centering 
\includegraphics[scale=1]{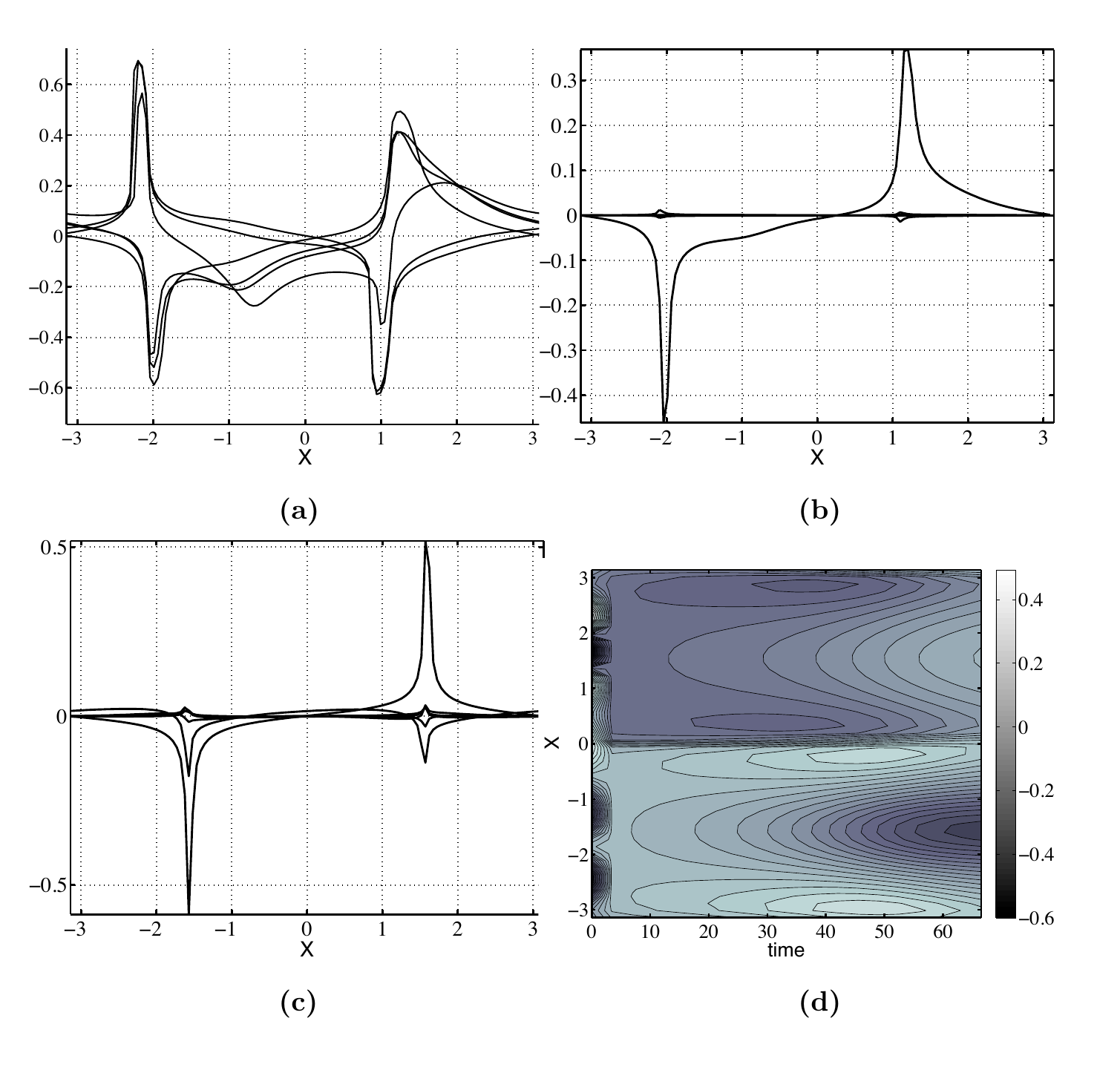}
\caption{{\it (a) At $t=70$, the superposition of 6 BV simulations, the $j^{th}$  one  corresponding to a sine wave perturbation, as per (\ref{sins}). 
(b) Cross section of the EBVs, at $t=70$, run with an ensemble of perturbations (see \ref{sins}). 
We note that, unlike the BV case in (a), the EBV has settled into a structure that clearly reflects the dominant EBV ensemble member.
(c) Cross section of the same EBVs, at an earlier time: at $t=5$. Even for short times, the EBV already shows an amplitude-ordering of its vectors. 
(d) Superposition of all EBVs, as a function of time and space.
}} 
\label{compert} 
\end{figure}

The shape or spectral content of the perturbation
mattered as well.
The spectrum of the perturbation is clearly important when projecting onto
spectral bases for a reduced representation. To illustrate this, 
we  use  the CY92 model, with the  same forcing as before and same
initial condition. We examine   monochromatic 
sine wave perturbations with wavenumber $j=1,2,...,6$ of the form
\begin{equation} 
\delta Y_0 = \epsilon_j  \sin [j x + \frac{1}{3} (j-1) \exp(1)],
\label{sins}
\end{equation}
where all of the perturbation amplitudes remain the same, $\epsilon_j=0.25$. The choice of the 
phase is inconsequential: to show this we chose non-commensurate phases among the
sine wave components.
In Figure \ref{compert}a we show  the BVs associated with each of the perturbations
($j=1,2...,6$), at $t=70$. 
Figure \ref{compert}b shows the EBVs at  $t=70$. Even at  $t=5$, shown
in Figure \ref{compert}c, we already see an amplitude-ordered
structure in the EBV. 

The space-time plot of all EBVs is plotted in Figure \ref{compert}d;
we note the very short transient phase, lasting till about $t=4$,
followed by a structure which is clearly dominated by the largest member of
the EBV. 
The EBV calculation was run with identical parameters to those used in Figure \ref{compert}a, with 
the ensemble consisting of  the same initial perturbations in
(\ref{sins}). In the BV case we are getting outcomes that do not have the
reductive appearance of the EBV. On the other hand, there is no ambiguity to the prevailing ensemble member in the EBV case.
This ensemble member has a clear correspondence in structure to  the path and its
perturbation field. The BVs should eventually agree with the EBVs, nevertheless. It is 
safe to assume  that this will happen only after a very long time, longer than the time interval that might 
be suggested by the base solution.

\begin{figure}
\centering 
\includegraphics[scale=0.7]{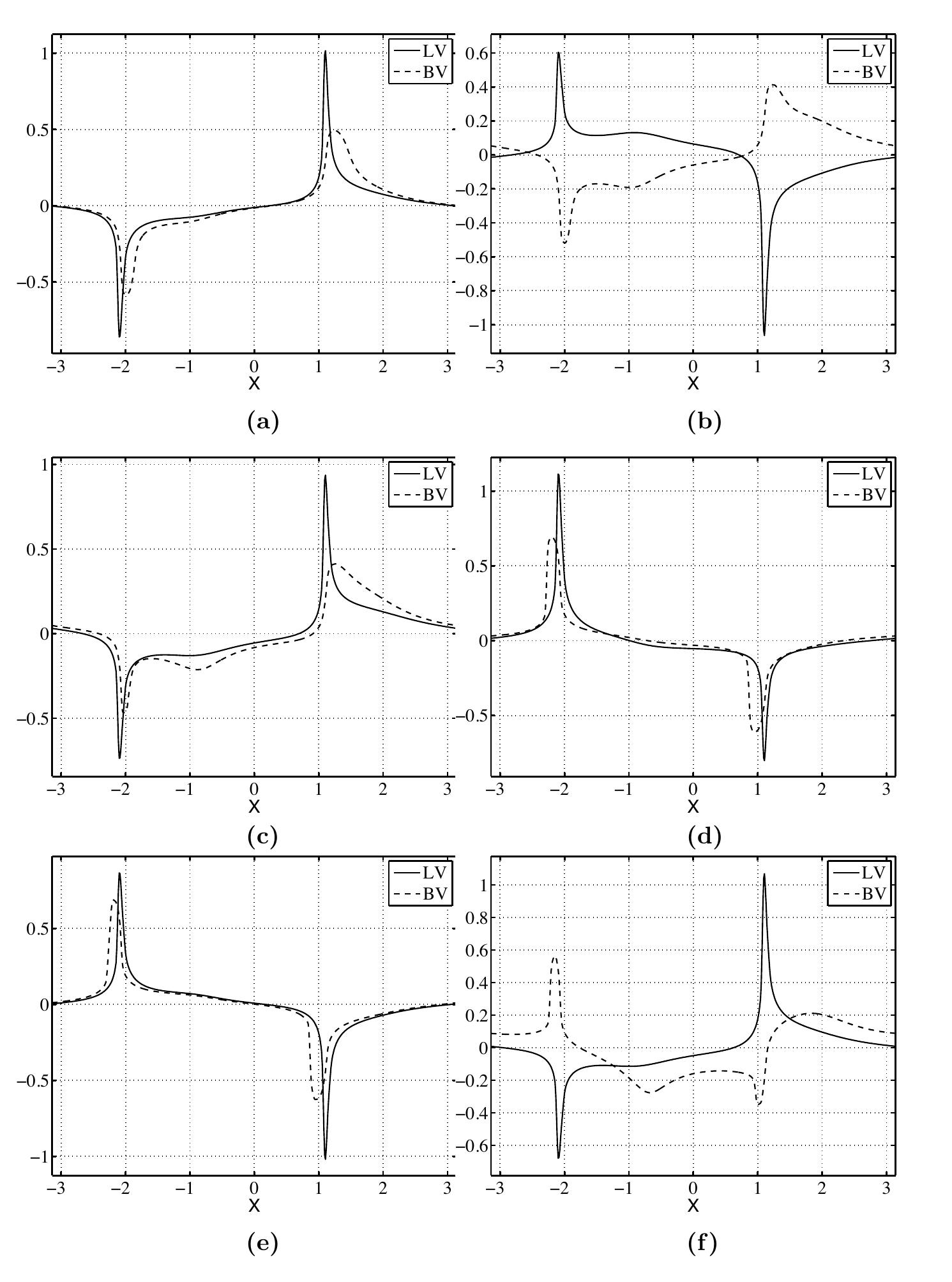} 
\caption{{\it Wavenumber dependence of BVs and finite-time Lyapunov vectors.  CY92 Model, cosine initial condition and non-symmetric forcing (see Figure \ref{cycase}).  Comparison of the finite-time Lyapunov vectors and the BVs for sine wave perturbations of size $\epsilon_j=0.25$. The perturbations are the individual sine waves in (\ref{sins}),  (a)  $j=1$, (b) $j=2$, ..., (f) $j=6$.}} 
\label{perts} 
\end{figure}

We also compared BV, for each of the perturbations in (\ref{sins}), to the outcomes of the finite-time Lyapunov vector calculation. 
Figure \ref{perts} compares the BVs and finite-time Lyapunov vectors for each $j$ wavenumber perturbation. The amplitudes are set to
$\epsilon_j=0.25$, for $j=1, 2, ...,6$.   The structure of the finite-time Lyapunov vectors, to within a sign, is qualitatively the same,
 regardless of the
wavenumber of the perturbation. The BVs are not. If the perturbation was made considerably smaller the differences
between the BVs and the finite-time Lyapunov vectors would become small, as expected:
As shown in Section \ref{interpret}, 
the BVs and the finite-time Lyapunov vectors must be similar to each other, provided the 
perturbations are small enough. For larger amplitudes, the BVs might
show more structure since nonlinear effects could
play a role in the structure of the BVs. Apparently, $\epsilon=0.25$ is 
already in the
range of large perturbations of the CY92 about the 
solution chosen. It was not clear whether a chosen perturbation is large
or small, based solely on the CY92 model itself: it was only clear to us after a comparison of the outcomes of the finite-time Lyapunov case and the BV case. 

\subsection{Cahn-Hilliard Dynamics}
\label{sec.CHD}

The Bi-harmonic Operator $B\,u\,=\,\triangle^2\,u$
on $\Omega$ and the linear Bi-harmonic equation: 
\begin{equation}
\label{bharm}
\partial_tu+\nu\triangle^2\,u\,=\,0
\end{equation}
play a basic role in the study of the CHE.
We assume for now that the Operator $B$ satisfies either the
non-flux  boundary conditions:
\beq
\label{homobc}
\frac{\partial u}{\partial n}\,=\,0
\quad\text{and}\quad
\frac{\partial(\triangle u)}{\partial n}\,=\,0,
\qquad\text{on }\Gamma,
\eeq
or the periodic boundary conditions, see \cite{SY02}.
In the sequel, we will assume that the forcing function $F$
in \eqref{CH1} satisfies $F\in H_0^j$, for some $j\geq0$.
That is to say, 
$$
\int_{\Omega}F\,dx\,=\,0.
$$
Thus, by integrating \eqref{CH1}, one observes that any solution 
$u=u(t,x)$ satisfies
\beq
\label{s-ave}
\overline{u}_0\, := \,\frac1{|\Omega|}\int_{\Omega}u_0(x)\,dx
\,=\,\frac1{|\Omega|}\int_{\Omega}u(t,x)\,=\,\overline{u}(t),
\qquad\text{for }t\geq0.
\eeq

\sep
One seeks solutions of the CHE \eqref{CH1}
in the Sobolev space $V_0^2$.
A {\em mild} solution $u=u(t)$ of the initial value problem is 
given by
\begin{equation}
\label{mild}
u(t)\,=\,e^{-\nu B_0 t}u_0+\int_0^te^{-\nu B_0(t-s)}\,\left[
	\triangle(g(u(s)))+F\right]\,ds.
\end{equation}
We will denote the maximally defined solutions of \eqref{mild}
by $u(t)\,=\,S(t)u_0$. With $u_0\in V_0^2$, this mild solution is
uniquely determined, with $S(t)u_0\in V_0^2$, for
$0\leq t<T(u_0)$, where $0<T(u_0)\leq\infty$. 

\sep
Because of \eqref{s-ave},
we see that, for $j\geq2$, the spaces
\begin{equation}
\label{H0j}
H_0^j\, := \,\{\phi\in H^j\,:\,\overline{\phi}=0\}
\end{equation}
are positively invariant spaces for the solutions of \eqref{CH1}.
Thus we will focus only on solutions $S(t)\,u_0$
that satisfy $\overline{u}_0=0$.
We also denote the collection of stationary solutions of the CHE by
$$
Q\, :=\,\{u_0\in H^2\,:\,S(t)u_0=u_0\text{ for all }t \in \BR\},
$$
and $Q_0=Q\cap H_0^2=Q\cap V_0^2$.

\subsubsection{The CHE With $F\equiv 0$:}

The basic problem with forcing $F\equiv0$ is of special interest.
The equation \eqref{CH1} becomes:
\begin{equation}
\label{CH0}
\partial_tu+\nu\triangle^2\,u\,=\,\triangle(g(u))
\qquad\text{for }x\in\Omega\text{ and }t\geq0.
\end{equation}
To study the solutions of \eqref{CH0}, one uses the
Landau-Ginsburg functional:
\beq
\label{lg}
J(u)\, := \,\int_\Omega\bigg[\frac{\nu}{2}|\nabla u|^2+G(u)\bigg]\,dx,
\qquad\text{where }G(z)=\int_0^z g(s)\,ds.
\eeq
In addition, the Landau-Ginsburg functional satisfies
\beq
\label{LL1}
\partial_t J(S(t)u_0)\,=\,-\|\nabla K(S(t)u_0)\|_0^2,
\qquad\text{for }0<t<T(u_0),
\eeq
where $K(u)=-\nu\triangle u+g(u)$. 
Furthermore, there exist positive constants
$f_0$ and $C_0$ such that $-f_0\leq G(s) \leq C_0s^{4}+f_0$ and
\beq
\label{LL2}
\frac{\nu}2\|\nabla S(t)u_0\|_0^2-f_0|\Omega|\leq J(u_0).
\eeq
Since $J(u_0)$ is bounded below, see \eqref{LL2},
and decreasing along orbits, see \eqref{LL1},
it follows that $J(S(t)u_0)$ is a Lyapunov function,
see \cite{LL61}.
This implies that the mild solution $S(t)u_0$ is defined for all
$t>0$, i.e., $T(u_0)=\infty$, for every $u_0\in V^2$.
It is fact that, whenever $F\equiv0$ and $\overline{u}_0=0$,
then $\omega(u_0)$, the omega limit set of the solution
$S(t)\,u_0$ is a nonempty, compact, connected
invariant set in $Q_0$.

Let us now return to the CHE with forcing \eqref{CH1}.
By integrating the equation \eqref{CH1}, where $u_0\in H_0^2$
and $F\in H_0^2$, as well, then $S(t)\,u_0\in H_0^2$,
for all $t\geq0$.

\subsubsection{The Global Attractor $\fma_0$.}

Assume that $B$ satisfies the boundary conditions BC, {\it i.e.},
either the non-flux condition
\eqref{homobc} holds, or the periodic boundary conditions hold.
Let $S(t)$ be the semiflow generated by the solution operator on $V_0^2$. 
 Then the following hold.
\begin{enumerate}
\item $S(t)$ has a nonempty, compact 
	global attractor $\fma_0$ in $V_0^2$,
	and $\fma_0$ attracts all bounded sets in $V_0^2$.
	The attractor $\fma_0$ depends continuously
	on the forcing function $F\in H_0^2$.
\item When $F\in H_0^3$, then the attractor $\fma_0$ is a compact, invariant set in $V^{4r}$,
	for each $r$ with $0\leq r\leq1$.
\item The set $Q_0$ is nonempty, compact, and invariant with
	$Q_0\subset\fma_0$.
\item Lastly, there is an Inertial Manifold for the solutions 
	of the infinite dimensional system \eqref{CH1}, 
	see \cite{FST} and \cite{SY02}. 
	One finds this manifold by using the
	orthogonal projection $\BP_N$ onto the lowest 
	$N$ nodes, that is to say, into 		
	$\text{Span}\{e_k\,:\,k\leq N\}$, where
	$\{e_k\}$ are the eigenfunctions for the Bi-harmonic
	operator $B$. One then makes a change of variables
	$u=v+w$, where $v=\BP_N\,u$, $w=\BQ_N\,u$, and
	$\BQ_N=I-\BP_N$. The $(v,w)$ system:
\begin{equation}
\label{BG}
\begin{aligned}
\partial_t v +Bv\,&=\,\BP_N\left(\triangle(g(v+w))+F\right) \\
\partial_t w +Bw\,&=\,\BQ_N\left(\triangle(g(v+w))+F\right), 
\end{aligned}
\end{equation}
	is equivalent to the CHE \eqref{CH1}.
	One then shows that, for $N$ large,
	the variable $w$ is enslaved to the variable $v$
	in some neighborhood of $\fma_0$.
	That is to say, $w\,=\,\Phi(v)$, for a suitable function $\Phi$.
	It turns out the the longtime dynamics of \eqref{CH1} 
	is equivalent to the longtime dynamics of the 
	finite-dimensional ordinary differential equation:
\begin{equation}
\label{IM}
\partial_t v +Bv\,=\,\BP_N\left(\triangle(g(v+\Phi(v)))+F\right).
\end{equation}
	See \cite{SY02} and the references contained therein, for more
	details.
\end{enumerate}

\subsection{BV and  EBV in the Higher Nonlinear Regime}

For sufficiently small perturbations and short renormalizing 
time intervals the finite-time Lyapunov vector, the BV algorithm, and the EBV are similar in outcomes. As we depart from these conditions, we see significant differences in the outcomes of the three methods. The BV outcomes are most
sensitive to the amplitude and the frequency of  the perturbations. Regardless of the amplitude of the perturbations, the EBV outcomes are structurally unambiguous and robust.

\begin{figure}
\centering
\includegraphics[scale=1]{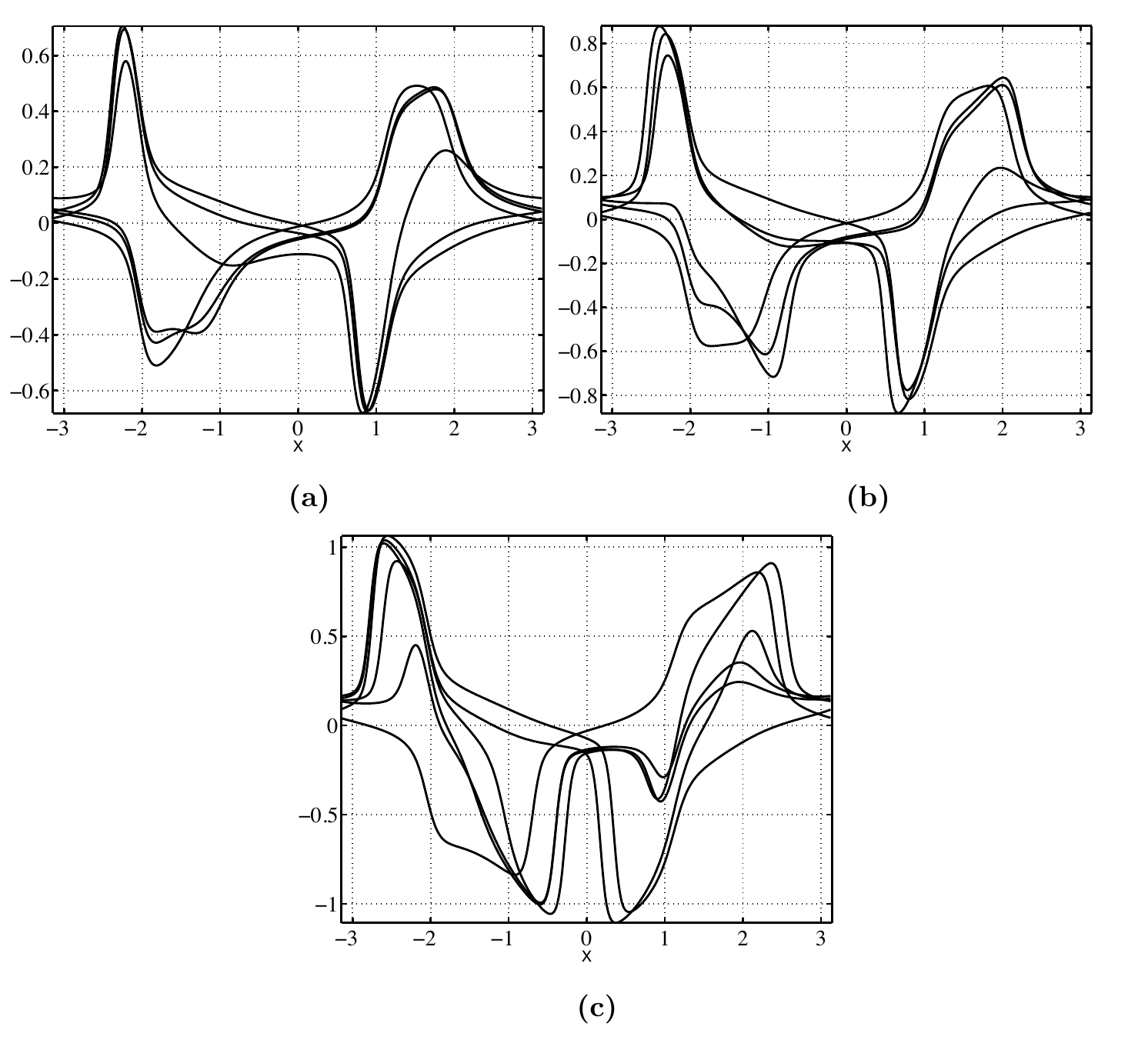} 
\caption{{\it BVs at $t=19.025$, corresponding to $\epsilon_j = \epsilon$, for all $j$.  (a) 
$\epsilon=0.6$, (b) $\epsilon=0.8$, (c)  $\epsilon=1.2$.
The outcomes are very sensitive to nonlinearity. There is a resulting ambiguity in structure of the perturbation field.}}
\label{nonlineareffectsBV}
\end{figure}

\begin{figure}
\centering
\includegraphics[scale=1]{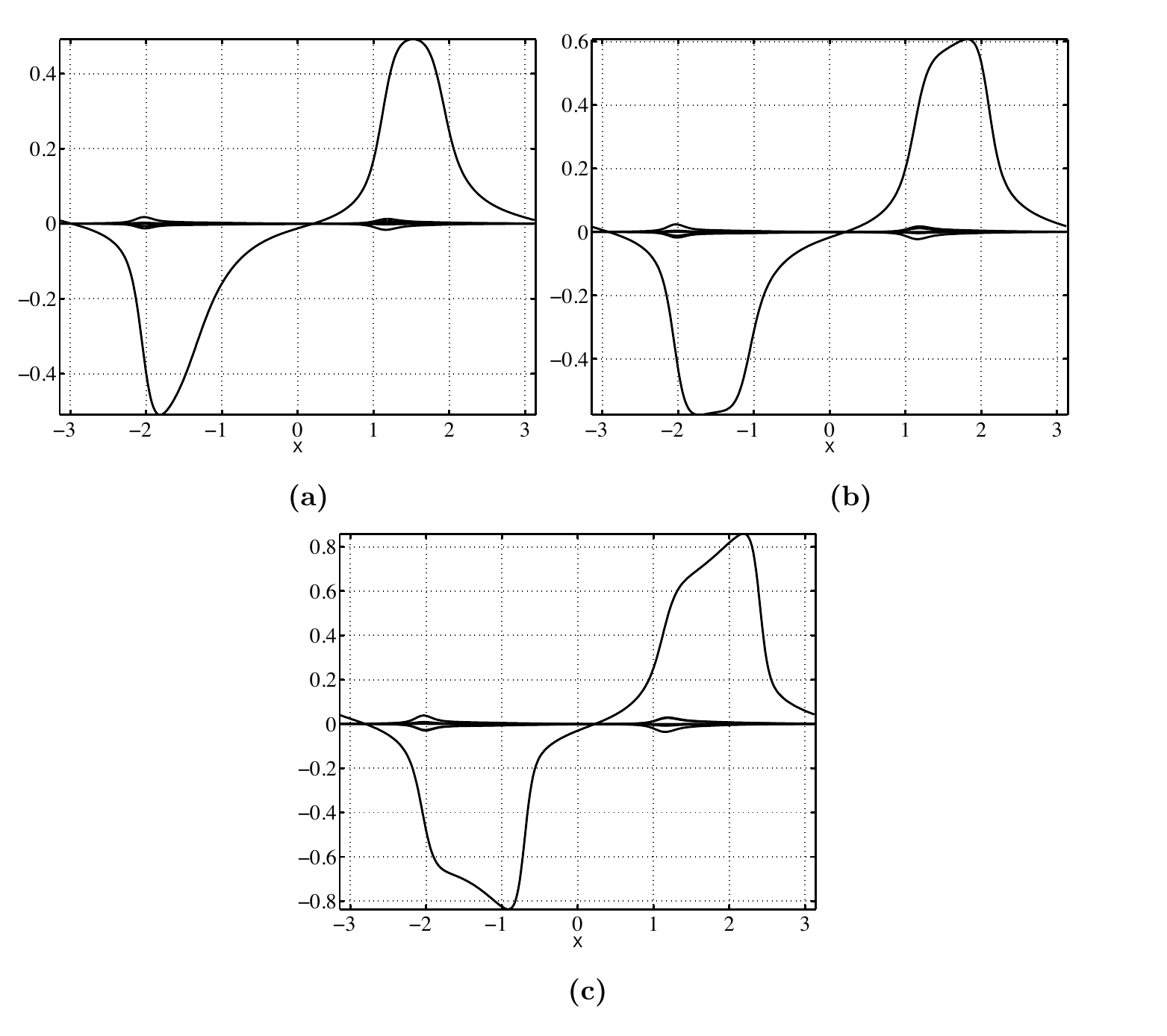} 
\caption{{\it EBV with  (a) $\epsilon=0.6$, (b) $\epsilon=0.8$, (c)  $\epsilon=1.2$. 
Compare outcomes to Figure \ref{nonlineareffectsBV}.
The vectors are shown in their original scales.}}
\label{nonlineareffectsEBV}
\end{figure}

\begin{figure}
\centering
\includegraphics[scale=1]{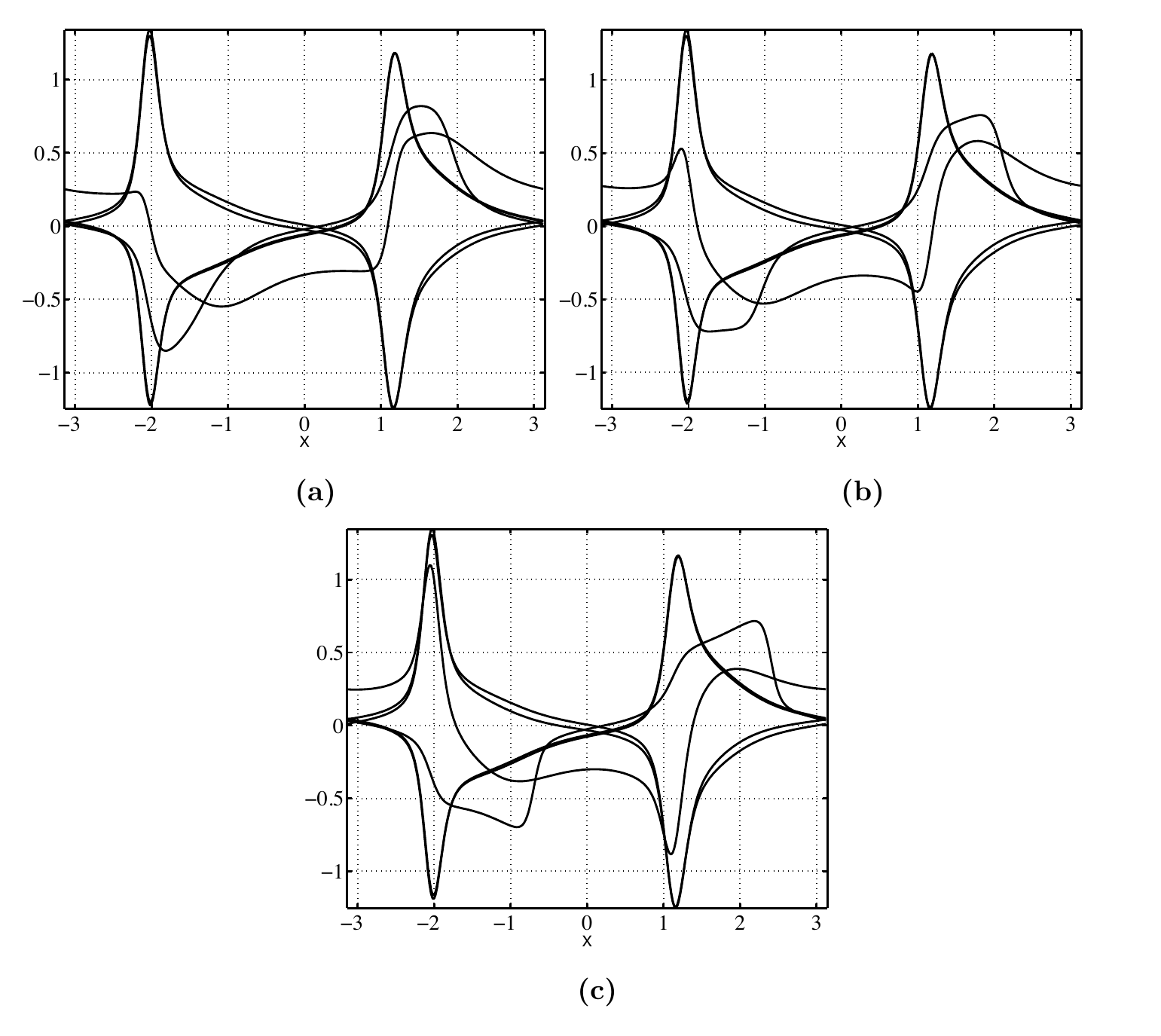} 
\caption{
{\it EBV outcomes shown in Figure \ref{nonlineareffectsEBV}, with 
each vector rescaled to an $L^2$-norm of 1 to aid in visual comparison.
Perturbation amplitudes (a) $\epsilon=0.6$, (b) $\epsilon=0.8$, and (c) $\epsilon=1.2$.
Note that as the perturbation amplitude increases, the resemblance between the BV and the EBV outcomes is lost, except for one of the vectors. }
}
\label{nonlineareffectsEBVSC}
\end{figure}

We  revisit the CY92 simulations of BV and EBV,  using the same forcing and initial  conditions used Section \ref{cy92} but  set the time scale parameter $\alpha=0.01$, we increase the spatial 
resolution to 320 points, and set the integration time step at $\delta t=0.000015$.  
The perturbation field is the same as in 
(\ref{sins}); however, we will increase the size of the perturbation of each of the components. 
For $\epsilon_j = 0.6, 0.8$, and $1.2$, with $j=1,2,...,6$, in (\ref{sins}) we obtain Figure \ref{nonlineareffectsBV}, 
which shows all BVs at $t=19.025$. We expect a simple structure in the perturbation field and thus the BV should
reflect this. The EBV results appear in Figure \ref{nonlineareffectsEBV}.
Comparison of Figures \ref{nonlineareffectsBV} and \ref{nonlineareffectsEBV}
show how the BV outcomes look qualitatively different from their EBV counterparts 
as we increase the initial amplitude of the ensemble.  The comparison is further
aided by reference to Figure \ref{nonlineareffectsEBVSC}, in which each ensemble
member of the EBV, at $t=19.025$,  has been rescaled to  1 in amplitude.
The BV outcomes shown in Figure \ref{nonlineareffectsBV} do not yield the structural clarity
that the EBV ensemble displays in Figure \ref{nonlineareffectsEBV}; this is a natural
consequence of the size-ordering inherent in the EBV algorithm.

In reference to Figure \ref{nonlineareffectsBV}c and Figure \ref{nonlineareffectsEBV}c
we see the similarity between one and only one of the BV vectors and the largest EBV.
A striking structural feature in the EBV results shown in Figure \ref{nonlineareffectsEBV} is that
there are only 3 significantly-different shapes, among all of the six
vectors (see Figure \ref{nonlineareffectsEBVSC}, where the vector have
been rescaled to have size $1$ in $L^2$.) For relative 
comparison, see Figure \ref{nonlineareffectsBV}.  

The sensitivity of the outcomes to the size of initial perturbations
is significantly different in the EBV and the BV outcomes. There are
shape variations among the BVs  as the 
amplitude of the perturbation is increased, as evidenced in Figure \ref{nonlineareffectsBV}. 
This degree of sensitivity is not as prevalent in the EBV outcomes, shown in Figure 
\ref{nonlineareffectsEBVSC}. The BV algorithm does not distinguish between different perturbation
scales. A large perturbation BV calculation will thus not yield information concerning smaller
scales. In contrast, see Figure \ref{nonlineareffectsEBVSC}c, for the EBV case, where 
small scale information is evident. (See also Figure \ref{nonlineareffectsEBV}c).

\section{Implementation Issues}
\label{conc}

Like the BV algorithm, the EBV algorithm is capable 
of dealing with legacy code. The important
difference between implementing a code that does an ensemble of BV and the EBV is that
while the former  can be run concurrently, in the EBV the ensemble members require normalization
to each other. In terms of coding, this is a minor issue, if the state variable dimension of the underlying 
model is moderate. For very large problems communication becomes an issue, but not at all unfamiliar
in concurrent or hybrid computing. Any additional computational issues borne by the EBV  are 
well outweighed by the higher informational content of the EBV over the BV. 
Moreover, the EBV is much more 
robust under the nonlinear effects, as illustrated in Section \ref{examples}, than BV.
This last aspect is very important and it should be studied in the context of more chaotic
equations in fluid mechanics, meteorology, and geophysics.

\begin{figure}
\begin{center}
\includegraphics[scale=1]{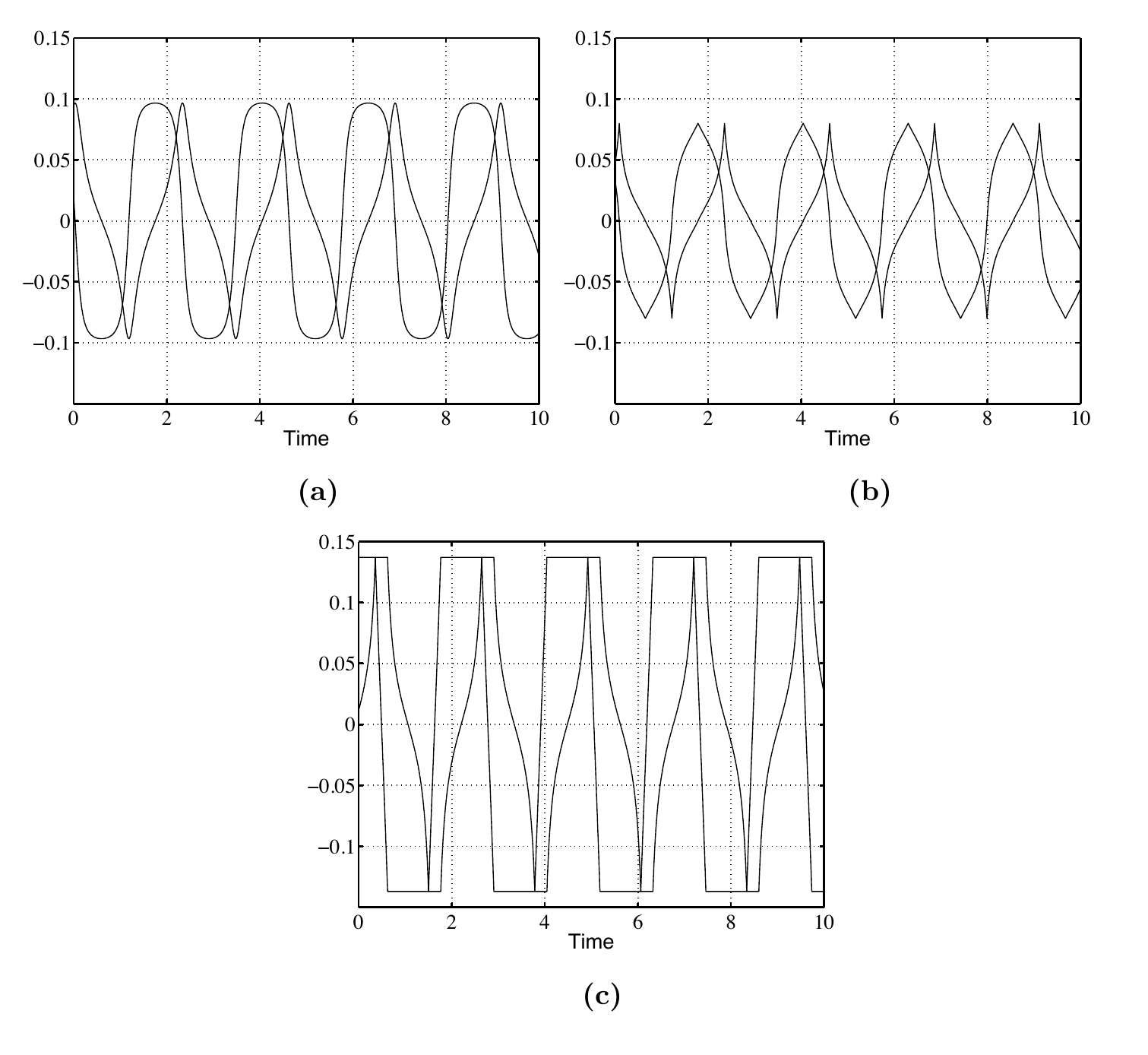} 
\end{center} 
\caption{{\it  (a) $L_2$-norm, (b) $L_1$-norm, and (c) $L_\infty$-norm of a sample BV as a function of time, for the  system 
in (\ref{sinsys}). The initial conditions are the same in all cases, the amplitude of the perturbations 
was $0.15$.}} 
\label{sinenorm}
\end{figure}

The numerical outcome of the BV and EBV algorithms depends on
the choice of norm used for rescaling. This is alluded-to in \cite{RiviereEtAl08}, but the reason
for this dependence turns out to be easily explained and can 
 be significant if  nonlinear effects are not negligible. The dependence of the outcomes on the norm 
 lies in the fact that it is not possible, in the general case,  to 
scale out the norm in the algorithm if the underlying dynamics   are nonlinear. To illustrate the norm dependence, we consider  a simple   2-dimensional system
\begin{eqnarray}
\frac{d X_1}{d t} &=& X_2, \nonumber \\
\frac{d X_2}{dt} &=&- \sin(8\, X_1), \quad t>0, \label{sinsys}
\end{eqnarray}
subject to the (same) initial conditions $(X_1(0),X_2(0)) =
(0.8,-1)$. Figure \ref{sinenorm} shows the outcomes of the BV algorithm for three
different choices of norms, starting from perturbations of size no greater than 
$0.15$. The norms we employ are all standard (finite) $L_p$ norms. The resulting BVs 
reflect the characters of these norms.
That  different outcomes are obtained is not surprising: the shape of
the unit sphere changes depending on which $L_p$ norm is used. 

A more weighty consideration related to norms concerns  
physical and  theoretical considerations.
Conservation laws provide guidance for the most appropriate norms for
the given dynamics (for instance, in Rayleigh-B\'ernard convection, the
temperature enjoys a maximum principle, while the velocity does not), 
but other considerations will play a role  ({\it e.g.}, a sup-norm
may be an obvious choice in determining the location of severe weather
events). It will not be
uncommon, thus, that a  mixture of norms may be necessary in multi-physics problems; the fact
that the choice of norm in the EBV/BV affects the results is in fact a good thing, not a bad one.

To end this section we wish to highlight a practical consideration
that may be not  be familiar
to practitioners, implementing BV or EBV algorithms computationally. The specific issue is
the impact of finite precision computing on the outcomes. It is easy to show that these algorithms
have high numerical sensitivity. In order to illustrate how this plays out we will consider the 
problem of calculating the vectors associated with 
\begin{equation}
\frac{dX}{dt} = A X, \quad t >0, \qquad X(0) = X_0. 
\label{aeq}
\end{equation}
Here $A$ is a square matrix of constants. For linear problems BV, EBV (and the finite-time Lyapunov)
algorithms must yield the same vectors. We will choose to illustrate computationally numerical 
ill-conditioning on a problem that exhibits transient growth due to the non-normal structure of the 
matrix $A$. We  emphasize, however, that the numerical sensitivity we will be highlighting 
in this example does not hinge on the nature of the dynamics, but rather, on the algorithmic 
form of the BV and EBV themselves.

We take 
$A$ to be an upper-triangular Jordan-block matrix of dimension 5, 
 where $A$ has a single eigenvalue $\lambda$, which is repeated
on the main diagonal, while the diagonal directly above the main
diagonal has non-zero entries. For simplicity we assume
these to  have the same value $\rho$. We assume further that 
$\lambda=-1$ and $\rho=1$.
The general solution operator contains ``generalized'' 
eigen-solutions with growth rates $t^k\,e^{-t}$, where 
$0\leq k\leq4$. (In this case, the eigenvalue has (algebraic)
multiplicity 5.)
  For this linear problem we expect to see, provided we take $t$ sufficiently  long to forget the transient,
 a very trivial outcome to BV, or EBV.

\begin{figure}
\begin{center}
\includegraphics[scale=1]{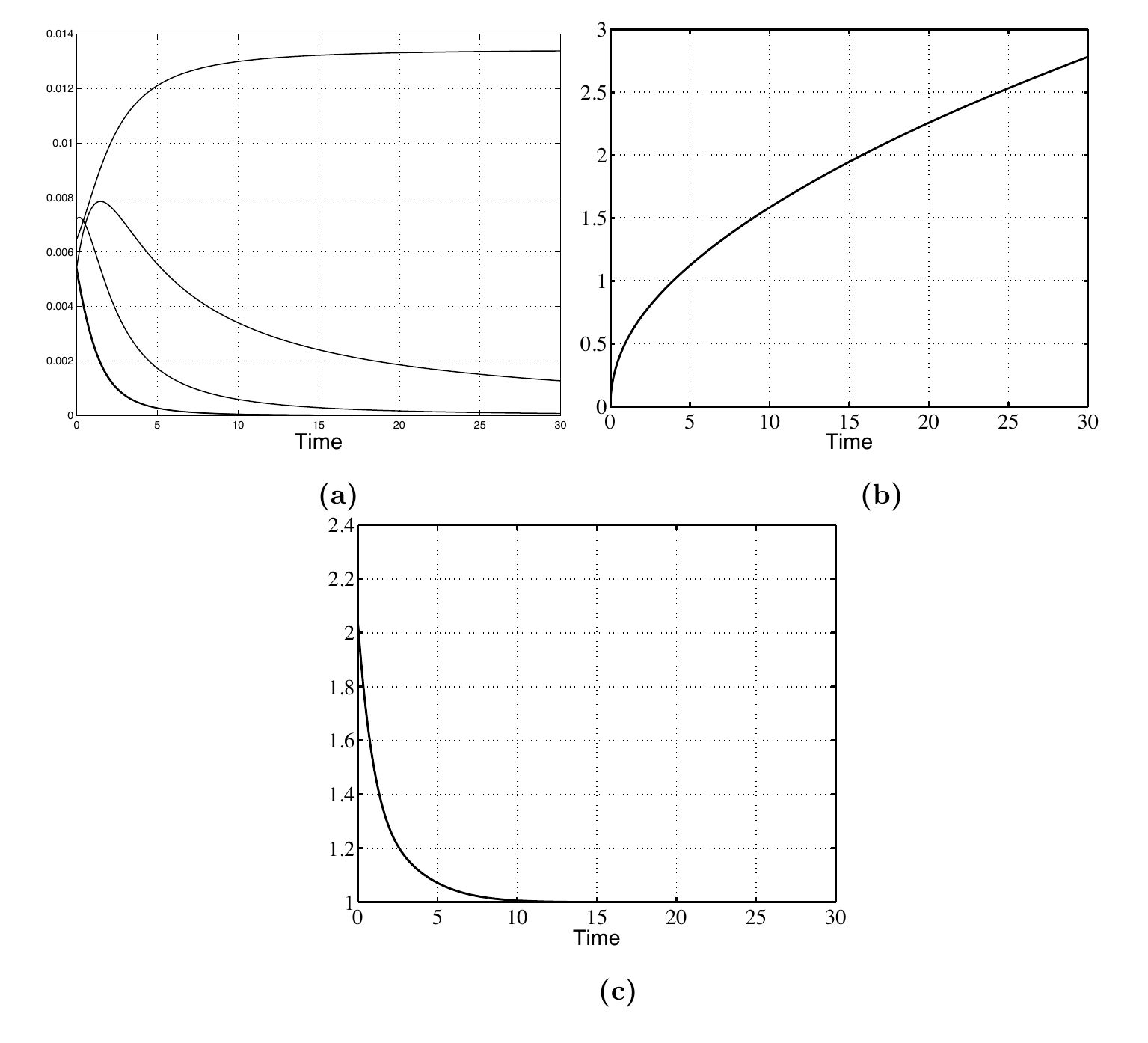} 
\end{center} 
\caption{{\it  (a) A sample BV, as a function of time. The calculation diverges and eventually 
fails (not shown); (b) 2-norm of the numerically-approximated $M(Y_n+\delta {\cal Y}_n,\delta t)-M(Y_n, \delta t)$ as a function of time, for the linear system  $\frac{dX}{dt} = A X$. 
 Here, $A$ is an upper-triangular Jordan-block matrix of dimension 5, 
 where $A$ has a single eigenvalue $-1$, which is repeated
on the main diagonal, while the superdiagonal has entries 1. (c) 2-norm of the numerically-approximated $M(Y_n+\delta {\cal Y}_n,\delta t)-M(Y_n,\delta t)$, well-conditioned case.  }} 
\label{ill}
\end{figure}

In Figure \ref{ill} we summarize  the results of the BV calculation on this 
system. 
We
  employed an integration time step of $0.001$, a perturbation initially of magnitude $0.01$.  We 
 performed  the calculation in double precision, using an explicit Runge-Kutta 4, but payed little
 attention to   how numerical sensitivity was handled. 
  In Figure
 \ref{ill}a we show the BV, {\it i.e.}, ${\cal Y}$, as a function of time. 
The  calculation of  BV, using
 (\ref{EQ3}),  clearly diverged, shortly after about $t=30$ (and thus not shown in Figure \ref{ill}a).
However, there were indications that something was not right even before it became obvious
that the solution was wrong: In Figure \ref{ill}b we show the 
 2-norm of $M(Y_n+\delta {\cal Y}_n,\delta t)-M(Y_n, \delta t)$. It  monotonically increases,  
even though all the factors $t^k\,e^{-t}$, for $0\leq k\leq 4$ in the solution
operator,  decay for $t>4$.

The outcome is not related to choosing a rescaling time to be too long: in fact, in this computation  
the rescaling is performed at each computational step. Moreover, 
the time step chosen was sufficiently small to guarantee asymptotic stability in the numerical
integrator. A smaller time step would have been able to ameliorate to a certain extent the 
numerical sensitivity of the difference $M(Y_n+\delta {\cal Y}_n,\delta t)-M(Y_n, \delta t)$, however, 
in very large scale problems this might not be practical or even possible.
Figure \ref{ill}c shows the 2-norm of  $M(Y_n+\delta {\cal Y}_n,\delta t)-M(Y_n,\delta t)$,
 using a numerically well-conditioned implementation of the BV  algorithm.  The
strategy  used to obtain a well-conditioned 
outcome was to normalize
the elements  in the required subtraction before computing their difference. 
The well-conditioned calculation was capable of qualitatively good
results. However, 
the  strategy adapted here to increase the numerical stability was by no means 
generally applicable to all problems, nor was it optimal.

\section{Concluding Remarks}

The main thrust of our work is to propose an ensemble-based vector breeding algorithm,
the Ensemble Bred Vector (EBV) algorithm. It is based on the  Bred
Vector (BV) algorithm introduced by  
\cite{tothkalnay93,tothkalnay97}. We compare the EBV to the BV algorithm and the finite-time Lyapunov Vector algorithms. 
In the EBV, an ensemble of initial perturbations is bred concurrently
and then rescaled by the size of the largest member of the bred
ensemble. The uniform normalization of all the ensemble members after each cycle is the
distinctive trait of the EBV algorithm, which leads to some profound differences when compared to the BV  algorithm.

As expected, when initial perturbations are sufficiently small, the EBV, the BV, and the finite-time Lyapunov vector algorithms lead to similar results. We gave a theoretical justification of this
phenomenon by looking at the corresponding time-continuum analogues of the BV and EBV algorithms. We rescale frequently: 
The algorithms and results are formulated, for simplicity, assuming that the rescaling is done after every discrete time step. 

In Section \ref{interpret}, we develop a solid mathematical basis for
both the BV and EBV algorithms and show that each
algorithm results in 
good approximations of the solutions of the tangent linear model,
when the step-size $\delta t$ is small. As is seen in Table \ref{tab1}, 
the EBV algorithm has a substantial
advantage over the (classical) BV algorithm, 
in the sense that the drop in the minimal EBV error is substantially
better than the corresponding drop in the minimal BV error.

In the study of the Lorenz attractor, the
classical BV algorithm has a shortcoming which limits one's ability
to use this algorithm to  study  the dynamics inside the
attractor.  In particular, equation \eqref{BVeq} implies that
$\|\delta \mathcal{Y}_{n+1}(\iota)\| = \epsilon$,
for all $n\geq0$ and all $\iota\in\mathbb{I}$.
The BV algorithm maps the cloud onto a 2-dimensional 
sphere. The third dimension is lost, and with it so are the
fractal patterns seen in Figures \ref{difflorenz2b} and \ref{zoom1}.
Also the many patterns seen in Figure \ref{zooma} are lost
because the BV alternative would map everything
onto the single line at height $\epsilon$.

Lastly, in Figures \ref{nonlineareffectsBV},  \ref{nonlineareffectsEBV}, 
and  \ref{nonlineareffectsEBVSC}, 
we examine a series
of related test problems that depart from the tangent linear model.
The point here is to get a comparison of the 
performance of the two algorithms, as one moves further
into nonlinear regime. Once again, one sees a significant
advantage of the EBV over the BV.

The theoretical aspects for the EBV and BV algorithms
are presented in Subsection 2.5.
As is noted there, our application to the basic issue of the
sensitivity with respect to errors in the initial conditions
relies heavily on the Johnson, et al manuscript, JPS87.

In conclusion, for 
the applications described in this article,
we believe that the new EBV algorithm has been shown to be
 superior to the traditional BV algorithm.
Finally, we ask: Is the EBV algorithm the "last" word
on modifications of the classical BV?
Probably not.
However, we do expect that the EBV will serve as a good
starting point for new theories of bred vectors.

\section*{Acknowledgements}
JMR was partially supported by NSF grant DMS-0335360. 
ALM was partially supported by NSF grant DMS-0708902, DMS-1009713 and DMS-1009714.
NB, ALM, JMR, and GRS  wish to thank
the Institute for Mathematics and its Applications (IMA) for their
support and hospitality. Research at the IMA is 
supported by the National Science Foundation and the University of Minnesota.
GRS acknowledges his appreciation of Robert Sacker for related suggestions he made at an early stage in
the development of this article.
We express our sincere gratitude to the referees, who made an in depth reading of the original manuscript 
and offered several helpful suggestions which led to improvements in the paper.

\vskip 4ex

\newpage


\begin{thebibliography}{33}
\providecommand{\natexlab}[1]{#1}
\providecommand{\url}[1]{\texttt{#1}}
\expandafter\ifx\csname urlstyle\endcsname\relax
  \providecommand{\doi}[1]{doi: #1}\else
  \providecommand{\doi}{doi: \begingroup \urlstyle{rm}\Url}\fi

\bibitem[Arnold(1998)]{larnold}
L.~Arnold.
\newblock \emph{Random Dynamical Systems}.
\newblock Springer, New York, 1998.

\bibitem[Buizza et~al.(1993)Buizza, Tribbia, Molteni, and Palmer]{btmp}
R.~Buizza, J.~Tribbia, F.~Molteni, and T.~Palmer.
\newblock Computation of optimal unstable structures for numerical weather
  prediction models.
\newblock \emph{Tellus}, 45A:\penalty0 388--407, 1993.

\bibitem[Cessi and Young(1992)]{cy92}
P.~Cessi and W.~R. Young.
\newblock Multiple equilibria in two-dimensional thermohaline circulation.
\newblock \emph{Journal of Fluid Mechanics}, 241:\penalty0 291--309, 1992.

\bibitem[Cheung(2001)]{Cheung01}
K.~K.~W. Cheung.
\newblock Ensemble forecasting of tropical cyclone motion: comparison between
  regional bred modes and random perturbations.
\newblock \emph{Meteorology and Atmospheric Physics}, 78:\penalty0 23--35,
  2001.

\bibitem[Corazza et~al.(2003)Corazza, Kalnay, Patil, Yang, Morss, Cai,
  Szunyogh, Hunt, and Yorke]{corazza03}
M.~Corazza, E.~Kalnay, D.~J. Patil, S.-C. Yang, R.~Morss, M.~Cai, I.~Szunyogh,
  B.~R. Hunt, and J.~A. Yorke.
\newblock Use of the breeding technique to estimate the structure of the
  analysis ï¿½errors of the dayï¿½.
\newblock \emph{Nonlinear Processes in Geophysics}, 10:\penalty0 233--243,
  2003.

\bibitem[Deremble et~al.(2009)Deremble, Dï¿½Andrea, and Ghil]{ddamg}
B.~Deremble, F.~Dï¿½Andrea, and M.~Ghil.
\newblock Fixed points, stable manifolds, weather regimes, and their
  predictability.
\newblock \emph{Chaos}, 19:\penalty0 043109, 2009.

\bibitem[Evans et~al.(2004)Evans, Bhatti, Kinney, Oann, Pe{\~n}a, Yang, and
  Kalnay]{evans04}
E.~Evans, N.~Bhatti, J.~Kinney, L.~Oann, M.~Pe{\~n}a, S.~Yang, and E.~Kalnay.
\newblock {RISE} undergraduates find that regime changes in {L}orenz« model are
  predictable.
\newblock \emph{Bulletin of the American Meteorological Society}, 85:\penalty0
  520--524, 2004.

\bibitem[Eyink(2005)]{cy92eyink}
G.~Eyink.
\newblock Statistical hydrodynamics of the thermohaline circulation in a two
  dimensional model.
\newblock \emph{Tellus, A}, 57:\penalty0 100--115, 2005.

\bibitem[Foias et~al.(1988)Foias, Sell, and Temam]{FST}
C.~Foias, G.~R. Sell, and R.~Temam.
\newblock Inertial manifolds for nonlinear evolutionary equations.
\newblock \emph{Journal of Differential Equations}, 73:\penalty0 309 -- 353,
  1988.

\bibitem[Gneiting and Raftery(2005)]{GneitRaft05}
T.~Gneiting and A.~E. Raftery.
\newblock Atmospheric science - weather forecasting with ensemble methods.
\newblock \emph{Science}, 310:\penalty0 248--249, 2005.

\bibitem[Hallerberg et~al.(2010)Hallerberg, Pazo, Lopez, and
  Rodriguez]{Hallerberg}
S.~Hallerberg, D.~Pazo, J.~M. Lopez, and M.~A. Rodriguez.
\newblock Logarithmic bred vectors in spatiotemporal chaos: Structure and
  growth.
\newblock \emph{Physical Review E}, 81, 2010.

\bibitem[Hansen and Smith(2000)]{HansenSmith00}
J.~A. Hansen and L.~A. Smith.
\newblock The role of operational constraints in selecting supplementary
  observations.
\newblock \emph{Journal of the Atmospheric Sciences}, 57:\penalty0 2859--2971,
  2000.

\bibitem[Johnson et~al.(1987)Johnson, Palmer, and Sell]{JPSell87}
R.~A. Johnson, K.~J. Palmer, and G.~R. Sell.
\newblock Ergodic properties of linear dynamical systems.
\newblock \emph{SIAM Journal of Mathematical Analysis}, 18:\penalty0 1--33,
  1987.

\bibitem[Kalnay(2003)]{kalnaybook}
E.~Kalnay.
\newblock \emph{Atmospheric Modeling, Data Assimilation and Predictability}.
\newblock Cambridge University Press, Cambridge, 2003.

\bibitem[LaSalle and Lefschetz(1961)]{LL61}
J.~P. LaSalle and S.~Lefschetz.
\newblock \emph{Stability by {L}yapunov's Direct Method with Applications}.
\newblock Academic Press, New York, 1961.

\bibitem[Lorenz(1963)]{lorenz63}
E.~N. Lorenz.
\newblock Deterministic nonperiodic flow.
\newblock \emph{Journal of Atmospheric Science}, 20:\penalty0 130--141, 1963.

\bibitem[Lyapunov(1992)]{AML92}
A.~M. Lyapunov.
\newblock The general problem of the stability of motion.
\newblock \emph{International Journal of Control}, 55\penalty0 (3):\penalty0
  521--790, 1992.
\newblock Translated by A. T. Fuller from {\'E}douard Davaux's French
  translation (1907) of the 1892 Russian original, 
  (historical introduction) by Fuller, a 
  Smirnov, and the bibliography 
  Barrett, 

\bibitem[Mandelbrot(1977)]{BM77}
B.~B. Mandelbrot.
\newblock \emph{Fractals: form, chance, and dimension}.
\newblock W. H. Freeman and Co., San Francisco, Calif., revised edition, 1977.
\newblock Translated from the French.

\bibitem[Mu and Jiang(2008)]{MuJiang08}
M.~Mu and Z.~N. Jiang.
\newblock A new approach to the generation of initial perturbations for
  ensemble prediction: Conditional nonlinear optimal perturbation.
\newblock \emph{Chinese Science Bulletin}, 53:\penalty0 2062--2068, 2008.

\bibitem[Palmer et~al.(1998)Palmer, Gelaro, Barkmeijer, and Buizza]{pgbb}
T.~N. Palmer, R.~Gelaro, J.~Barkmeijer, and R.~Buizza.
\newblock Singular vectors, metrics and adaptive observations.
\newblock \emph{Journal of Atmospheric Sciences}, 55:\penalty0 633--653, 1998.

\bibitem[Pliss and Sell(1999)]{plsell}
V.~A. Pliss and G.~R. Sell.
\newblock Robustness of exponential dichotomies in infinite-dimensional
  dynamical systems.
\newblock \emph{Journal of Dynamics and Differential Equations}, 11:\penalty0
  471--513, 1999.

\bibitem[Primo et~al.(2008)Primo, Rodriguez, and Gutierrez]{PrimoEtAl08}
C.~Primo, M.~A. Rodriguez, and J.~M. Gutierrez.
\newblock Logarithmic bred vectors. a new ensemble method with adjustable
  spread and calibration time.
\newblock \emph{Journal of Geophysical Research-Atmospheres}, page D05116,
  2008.

\bibitem[Rivi\'ere et~al.(2008)Rivi\'ere, Lapeyre, and
  Talagrand]{RiviereEtAl08}
O.~Rivi\'ere, G.~Lapeyre, and O.~Talagrand.
\newblock Nonlinear generalization of singular vectors: Behavior in a
  baroclinic unstable flow.
\newblock \emph{Journal of the Atmospheric Sciences}, 65:\penalty0 1896--1911,
  2008.

\bibitem[Sacker and Sell(1976)]{SS176}
R.~J. Sacker and G.~R. Sell.
\newblock Existence of dichotomies and invariant splittings for linear
  differential systems. {II}.
\newblock \emph{Journal of Differential Equations}, 22\penalty0 (2):\penalty0
  478--496, 1976.

\bibitem[Sacker and Sell(1977)]{SS77}
R.~J. Sacker and G.~R. Sell.
\newblock Lifting properties in skew-product flows with applications to
  differential equations.
\newblock \emph{Memoirs of the American Mathematical Society}, 11\penalty0
  (190):\penalty0 iv+67, 1977.

\bibitem[Sacker and Sell(1978)]{SS78}
R.~J. Sacker and G.~R. Sell.
\newblock A spectral theory for linear differential systems.
\newblock \emph{Journal of Differential Equations}, 27\penalty0 (3):\penalty0
  320 -- 358, 1978.

\bibitem[Sacker and Sell(1980)]{SS80}
R.~J. Sacker and G.~R. Sell.
\newblock The spectrum of an invariant submanifold.
\newblock \emph{Journal of Differential Equations}, 38\penalty0 (2):\penalty0
  135 -- 160, 1980.

\bibitem[Sell and You(2002)]{SY02}
G.~R. Sell and Y.~You.
\newblock \emph{Dynamics of Evolutionary Equations}.
\newblock Springer, New York, 2002.

\bibitem[Toth and Kalnay(1993)]{tothkalnay93}
Z.~Toth and E.~Kalnay.
\newblock Ensemble forecasting at {NCEP}: the generation of perturbations.
\newblock \emph{Bulletin of the American Meteorological Society}, 74:\penalty0
  2317--2330, 1993.

\bibitem[Toth and Kalnay(1997)]{tothkalnay97}
Z.~Toth and E.~Kalnay.
\newblock Ensemble forecasting at {NCEP}: the breeding method.
\newblock \emph{Monthly Weather Review}, 125:\penalty0 3297--3318, 1997.

\bibitem[Wang and Bishop(2003)]{WangBishop03}
X.~G. Wang and C.~H. Bishop.
\newblock A comparison of breeding and ensemble transform {K}alman filter
  ensemble forecast schemes.
\newblock \emph{Journal of the Atmospheric Sciences}, 60:\penalty0 1140--1158,
  2003.

\bibitem[Wei and Toth(2003)]{WeiToth03}
M.~Z. Wei and Z.~Toth.
\newblock A new measure of ensemble performance: Perturbation versus error
  correlation analysis {(PECA)}.
\newblock \emph{Monthly weather Review}, 131:\penalty0 1549--1565, 2003.

\bibitem[Wolfe and Samelson(2007)]{WolfeSamelson07}
C.~L. Wolfe and R.~M. Samelson.
\newblock An efficient method for recovering {L}yapunov vectors from singular
  vectors.
\newblock \emph{Tellus}, 59A:\penalty0 355--366, 2007.

\end{thebibliography}

\end{document}